\documentclass[12pt,a4paper]{article}
\usepackage{jheppub}
\usepackage[FIGTOPCAP]{subfigure}
\usepackage{amsmath}
\usepackage{amsfonts}
\usepackage{amssymb}
\usepackage{tensor,epigraph}
\usepackage{cancel}
\usepackage{ulem} 
\usepackage{musixtex}
\usepackage{booktabs}
\usepackage{empheq}
\usepackage{amsmath}
\usepackage{amssymb}
\usepackage{amsthm}
\usepackage{comment}
\usepackage{psfrag}
\usepackage{graphicx}
\usepackage{color}
\usepackage{bbm}
\usepackage{color}

\newcommand{\diff}[3][{}]{\frac{\mathrm{d}^{#1}{#2}}{\mathrm{d}{#3}^{#1}}}

\definecolor{klgreen}{rgb}{0.0, 0.5, 0.0}

\newcommand{\exclude}[1]{}

\newcommand{\beq}{\begin{equation}}
\newcommand{\eeq}{\end{equation}}
\newcommand{\bea}{\begin{eqnarray}}
\newcommand{\eea}{\end{eqnarray}}
\long\def\/*#1*/{}

\newcommand{\junk}[1]{}

\title{What's inside a hairy black hole in massive gravity?
}
\author[a]{Seyed Ali Hosseini Mansoori}
\author[b]{, Li Li}
\author[a]{, Morteza Rafiee}
\author[c]{, Matteo Baggioli}

\affiliation[a]{Faculty of Physics, Shahrood University of Technology, P.O. Box 3619995161, Shahrood, Iran\vspace{0.1cm}}
\affiliation[b]{CAS Key Laboratory of Theoretical Physics, Institute of Theoretical Physics,
Chinese Academy of Sciences, Beijing 100190, China \& School of Physical Sciences, University of Chinese Academy of Sciences, No.19A Yuquan Road, Beijing 100049, China \& School of Fundamental Physics and Mathematical Sciences, Hangzhou Institute for Advanced Study, UCAS, Hangzhou 310024, China.\vspace{0.1cm}}
\affiliation[c]{Wilczek Quantum Center, School of Physics and Astronomy, Shanghai Jiao Tong University, Shanghai 200240, China \& Shanghai Research Center for Quantum Sciences, Shanghai 201315.\vspace{0.1cm}}
\vspace{0.1cm}

\emailAdd{shosseini@shahroodut.ac.ir}
\emailAdd{liliphy@itp.ac.cn}
\emailAdd{m.rafiee@shahroodut.ac.ir}
\emailAdd{b.matteo@sjtu.edu.cn}
\abstract{}

\abstract{In the context of massive gravity theories, we study holographic flows driven by a relevant scalar operator and interpolating between a UV 3-dimensional CFT and a \color{black} trans-IR \color{black} Kasner universe. For a large class of scalar potentials, the Cauchy horizon never forms in presence of a non-trivial scalar hair, although, in absence of it, the black hole solution has an inner horizon due to the finite graviton mass. We show that the instability of the Cauchy horizon triggered by the scalar field is associated to a rapid collapse of the Einstein-Rosen bridge. The corresponding flows run smoothly through the event horizon and at late times end in a spacelike singularity at which the asymptotic geometry takes a general Kasner form dominated by the scalar hair kinetic term. Interestingly, we discover deviations from the simple Kasner universe whenever the potential terms become larger than the kinetic one. Finally, we study the effects of the scalar deformation and the graviton mass on the Kasner singularity exponents and show the relationship between the Kasner exponents and the entanglement and butterfly velocities probing the black hole dynamics. \color{black}Differently from the holographic superconductor case, we can prove explicitly that Josephson oscillations in the interior of the BH are absent.\color{black}}

\begin{document}
\maketitle
\section{Introduction}

Understanding the interior of a black hole is an intriguing and fundamental problem from both theoretical and experimental perspectives. Due to the highly non-linear nature of the Einstein's equations, it is typically difficult to obtain black hole solutions analytically. As a matter of fact, the structure and dynamics of the black hole interior, in particular, the region near the black hole singularity where the spacetime curvature becomes infinite, are still elusive concepts. Nevertheless, some analytical black hole solutions were found in the past. The first case is the neutral Schwarzschild black hole whose geometry displays an event horizon and a spacelike singularity within it~\cite{Schwarzschild:1916uq}. Other examples include Reissner- Nordstrom (RN)~\cite{https://doi.org/10.1002/andp.19163550905,1918KNAB...20.1238N} and Kerr black holes ~\cite{PhysRevLett.11.237}, corresponding to the cases with non-trivial electric charge and angular momentum, respectively. Both have an additional inner Cauchy horizon that represents a breakdown of predictability in general relativity and present a timelike singularity, appearing to violate the strong cosmic censorship (SCC) conjecture ~\cite{1969NCimR...1..252P}. More recently, in the framework of General Relativity, it has been shown that there is no Cauchy horizon for some kind of black holes with scalar hairs and symmetric horizons~\cite{Hartnoll:2020rwq,Cai:2020wrp,Hartnoll:2020fhc,Devecioglu:2021xug,An:2021plu,Grandi:2021ajl}. Under quite general conditions, the authors of~\cite{Yang:2021civ} showed that the number of horizons is highly constrained by classical matter.\\

On the other side, Holography~\cite{Maldacena:1997re} provides some important and promising probes of the black hole interior, such as correlation functions~\cite{Fidkowski:2003nf,Festuccia:2005pi}, entanglement entropy~\cite{Hartman:2013qma} and complexity~\cite{Stanford:2014jda,Brown:2015bva}. 
Following this line of investigation, recently the authors of~\cite{Frenkel:2020ysx} considered a
deformation of a thermal CFT state by a relevant scalar operator, which in the bulk yields a deformation of the Schwarzschild singularity, at late interior times, into a more general Kasner form.\,\footnote{The authors of~\cite{Frenkel:2020ysx} studied a free scalar minimally coupled to Einstein gravity with a negative cosmological constant. The generalization to the case with a scalar self-interaction term was studied in~\cite{Wang:2020nkd}.} This provides a first step beyond the non-generic and classically unstable black hole interiors which have been the main focus of previous holographic literature. At this point, it is imperative to extend such study to more general cases and to uncover possible novel features. One interesting case is that of massive gravity~\cite{Hinterbichler:2011tt,Rubakov:2008nh,Dubovsky:2004sg} in which the standard general relativity framework is modified by endowing the graviton with a nonzero mass. From the holographic perspective, massive gravity and the consequent (partial or not) breaking of diffeomorphisms invariance, realizes in a simple and effective way (in fact retaining the homogeneity of the background geometry) the breaking of translational invariance in the dual field theory~\cite{Vegh:2013sk}. In general, it is very helpful to write down the massive gravity theory in the so-called Stueckelberg form ~\cite{Baggioli:2014roa,Alberte:2015isw} where it appears as a simple set of shift-invariant massless scalar coupled to canonical massless gravity. The usages of this class of models in the Holographic community are very vast, specially in view of the applications to strongly coupled matter~\cite{Alberte:2016xja,Alberte:2017cch,Alberte:2017oqx,Baggioli:2020edn}. We refer the Reader to~\cite{Baggioli:2021xuv} for a complete and exhaustive review on the topic.\\

In the present work, in the context of the holographic massive gravity framework of ~\cite{Baggioli:2014roa,Alberte:2015isw}, we study holographic renormalization group flows induced by a neutral scalar field $\phi$ and interpolating between a 3-dimensional UV CFT and a singular Kasner-like universe in the \color{black} trans-IR \color{black}. In contrast to the previous cases~\cite{Frenkel:2020ysx,Wang:2020nkd} where the hairless background is given by the Schwarzschild solution lacking the Cauchy horizon, in the massive gravity case the black hole solution in the absence of the scalar hair $\phi$ can have an inner horizon due to the mass of graviton, which in some sense acts as a charge for the black hole\footnote{See for example the simplest model in~\cite{Andrade:2013gsa}.}.
We consider a relevant deformation of the dual CFT thermal state leading to a holographic renormalization group (RG) flow at finite temperature. Although in our framework, in absence of scalar hair, the black hole presents an inner Cauchy due to the finite graviton mass, the deformation induced by a neutral scalar operator generically removes this Cauchy horizon such that the deformed black hole with non-trivial scalar hair approaches a spacelike singularity at late interior time. The instability of the Cauchy horizon triggered by the scalar field leads to a rapid collapse of the Einstein-Rosen bridge at the would-be Cauchy horizon. The asymptotic geometry near the spacelike singularity is shown to take a general Kasner form whenever the kinetic term of the scalar field $\phi$ dominates. On the contrary, when the potential terms for $\phi$ become important, we find novel and interesting deviations from the standard Kasner universe. \color{black}Additionally, we prove analytically that, contrarily to the holographic superconductor case of~\cite{Hartnoll:2020fhc}, no Josephson oscillations are present in our case. Our findings suggest that such oscillations do not appear in the absence of background charge and a non-trivial bulk gauge field.\color{black}\\

Given these general facts, we also present a more detailed analysis of the black hole interior geometry in function of the various parameters of the model such as the graviton mass. Finally, we probe the aforementioned black hole interior by computing the entanglement velocity and butterfly velocity for the deformed black holes displaying a Kasner singularity at late times.\\

The rest of the paper is organized as follows. In Section~\ref{sec:setup}, we introduce the gravitational model used in this work, a general four dimensional massive gravity theory coupled to a neutral scalar field $\phi$. We present analytic black hole solutions without the scalar hair and prove the absence of a Cauchy-horizon for the hairy black holes. In Section~\ref{sec:ER} we discuss the collapse of the Einstein-Rosen bridge associated with the instability of the inner Cauchy horizon triggered by the scalar field. In Section~\ref{sec:kasner}, we construct the holographic flows from the AdS boundary to the Kasner singularity sourced by the scalar field. \color{black} In Section~\ref{newsec}, we prove analytically the absence of Josephson oscillations in our holographic model. \color{black} The violation of the Kasner form near the spacelike singularity is also discussed. The probes of the Kasner exponent are considered in Section~\ref{sec:probe}. We conclude with a brief discussion and some remarks for the future in Section~\ref{sec:discussion}.


\section{Holographic setup}\label{sec:setup}
We consider the following 4-dimensional bulk action
\begin{equation}\label{eq6}
S=\frac{1}{16\pi G} \int dx^4 \sqrt{-g}\Big[(R-2\Lambda)- K(X)-\partial^{\mu} \phi \partial_{\mu} \phi -V( \phi^2)\Big]\,,
\end{equation}  
where $G$ is the Newton constant, $R$ the Ricci scalar, $\Lambda$ the cosmological constant and $\phi$ a neutral bulk scalar field whose potential is denoted as $V$.
Additionally, we have introduced a set of shift-invariant massless scalar fields $\Phi^I (I=x,y)$ via their kinetic term $X\equiv\frac{1}{2} \sum_{I=1}^{2}\partial_{\mu} \Phi^{I} \partial^{\mu} \Phi^{I}$. Here $K$ is a generic scalar function \cite{Baggioli:2014roa,Baggioli:2016rdj} which is sometimes labelled as \textit{K-essence} \cite{Baggioli:2021ejg}. The bulk solution for the axion fields $\Phi^I$ is taken to be
\begin{equation}
\Phi^I=\alpha\, x^I,
\end{equation}
with $\alpha$ a constant. In this sense, these scalars break translational invariance and they provide a mass for the graviton \cite{Hinterbichler:2011tt,Rubakov:2008nh,Dubovsky:2004sg}; they are indeed nothing else that the Stueckelberg fields which do restore diffeomorphism invariance in our massive gravity theory.
In addition, the rest of the solution is parametrized as
\begin{equation}\label{metric1}
ds^2=\frac{1}{r^2} \left(-f(r) e^{-\chi (r)} dt^2+\frac{dr^2}{f(r)}+dx^2+dy^2\right),\quad \phi=\phi(r)\,,
\end{equation}
where, in our coordinate system, the AdS boundary is at $r = 0$ and the black hole singularity locates at $r \to \infty$.  Moreover, at the event horizon $r_+$, the blackening function $f(r_+)$ vanishes. From \eqref{eq6}, the independent equations of motion read
\begin{eqnarray}
r^4 e^{\frac{\chi}{2}} \left(\frac{e^{-\frac{\chi}{2}} f \phi'}{r^2}\right)'&=&\diff{ V}{\phi^2} \phi\,, \label{eq2}\\
\chi'&=&r(\phi')^2\,,\label{eq4}\\
4 e^{\frac{\chi}{2}} r^4 \left(\frac{e^{-\frac{\chi}{2}}}{r^3} f\right)'&=&2 V(\phi^2 )+2  K(\alpha^2 r^2)-12\,,\label{eq5} 
\end{eqnarray}
with a prime denoting the derivative with respect to $r$. We have also chosen the cosmological constant $\Lambda=-3/L^2$ with the AdS radius $L$ set to one. In general, these coupled differential equations do not allow analytical solutions, thus one has to solve them numerically.

Depending on the choice of the potential $K(X)$, the model in Eq.~\eqref{eq6} corresponds to different boundary field theories. In particular, the shape of the potential and its behaviour near the AdS boundary at $r=0$ determines whether translational invariance is broken explicitly or spontaneously in the dual field theory picture~\cite{Baggioli:2021xuv,Ammon:2019wci}.
In the present work, we are interested in the following two cases:
\begin{itemize}
\item[•] Type I: $K(X)=X$, which corresponds to the well-known linear axion model ~\cite{Andrade:2013gsa}.

\item[•] Type II: $K(X)=a_{1} \sqrt{X}+ a_{2} X$ with $a_1$ and $a_2$ two constants. This form of potential corresponds to the non-linear dRGT massive gravity model~\cite{Vegh:2013sk} as proven in~\cite{Alberte:2015isw}.
\end{itemize}
We stress that $K'(X)>0$ to avoid ghost instability~\cite{Alberte_2018}. Both type I and type II corresponds to the explicit breaking of translations in the dual field theory and their physics is that described by momentum dissipation.\\

To continue, near the AdS boundary $r\to 0$, the asymptotic expansion for the various fields $\{\phi, \chi,
f\}$ is, respectively, given as follows:
\begin{eqnarray}
	\phi&=&\phi_{0}\, r+ \left\langle O\right\rangle r^2+...\,,\\
	\chi&=&\frac{\phi_{0}^{2}}{2}\, r^2+\frac{4 \phi_{0} \left\langle O\right\rangle}{3} r^3+...\,,\\	 
   f&=&	\begin{cases}
		1-\frac{\alpha^2}{2} r^2+\phi_{0}^2 r^2-\left\langle T_{tt}\right\rangle r^3+...\,, \hspace{21mm} \text{for type I} \\
		1-\frac{\alpha\, a_{1}}{4} r -\frac{\alpha^2 \,a_{2}}{2} r^2 +\phi_{0}^2 r^2-\left\langle T_{tt}\right\rangle r^3+...\,, \hspace{5mm} \text{for type II}
	\end{cases}
\end{eqnarray}
where we have considered the scalar potential $V(\phi^2)=m^2 \phi^2$  with $m^2=-2$, and have taken the normalization of the time coordinate at the boundary such that $\chi(r=0)=0$. Here, $\phi_{0}$ is the source of the scalar operator of the boundary field theory and $\left\langle O\right\rangle$ the corresponding expectation value.\,\footnote{It should be noted that the mass square $m^2$ of the bulk scalar field $\phi$ determines the scaling dimension $\Delta$ of the dual operator $O$ according to
\begin{equation}
	\Delta=\frac{3}{2}+\sqrt{\frac{9}{4}+m^{2}}\,,\nonumber
\end{equation}
A negative value of $m^2<0$ corresponds to a relevant operator with $\Delta <3 $ in the three dimensional boundary theory. Throughout the rest of the paper, we will always consider standard quantization for all our bulk fields.} $\left\langle T_{tt}\right\rangle$ is the energy density of the thermal state in the boundary field theory, and the corresponding temperature reads
\begin{equation}
	T=-\frac{f' e^{-\frac{\chi}{2}}}{4 \pi}\Big{|}_{r=r_{+}}\,.
\end{equation}
After imposing regularity at the horizon, $r=r_{+}$, one can write down $\left\langle O\right\rangle$ and $\left\langle T_{tt}\right\rangle$ in terms of $T$, $\phi_{0}$ and $\alpha$. Nevertheless, our system enjoys the following scaling symmetry 
\begin{equation}
r\rightarrow r/\lambda,\quad (\phi_0, T,\alpha)\rightarrow \lambda(\phi_0,T, \alpha),\quad \left\langle O\right\rangle\rightarrow \lambda^2 \left\langle O\right\rangle,\quad \left\langle T_{tt}\right\rangle\rightarrow \lambda^3 \left\langle T_{tt}\right\rangle\,,
\end{equation}
with $\lambda$ a constant parameter. Therefore, without loss of generality, we shall work with dimensionless quantities in units of $T$. All in all, the final full solution is then parametrized solely by the dimensionless combinations $\phi_{0}/T$ and $\alpha/T$. For convenience, we fix $r_+=1$ in all the manuscript.

\subsection{Black holes with no scalar hair}\label{sec:hairlessBH}
The black hole solutions of~\eqref{eq2}-\eqref{eq5} in the absence of the scalar $\phi$ can be obtained analytically (see \cite{Baggioli:2014roa} for the general solution). It is clear that the only non-trivial equation is~\eqref{eq5} when $\phi=0$.\\

For the linear axion model (type I), from~\eqref{eq5}, we obtain
\begin{equation}\label{massivebh}
f(r)=1- \frac{\alpha^2 r^2}{2}+\frac{r^3}{r_{+}}\left(\frac{\alpha^2}{2}-\frac{1}{r_{+}^2}\right)\,.
\end{equation}
\begin{figure}
	\includegraphics[scale=0.6]{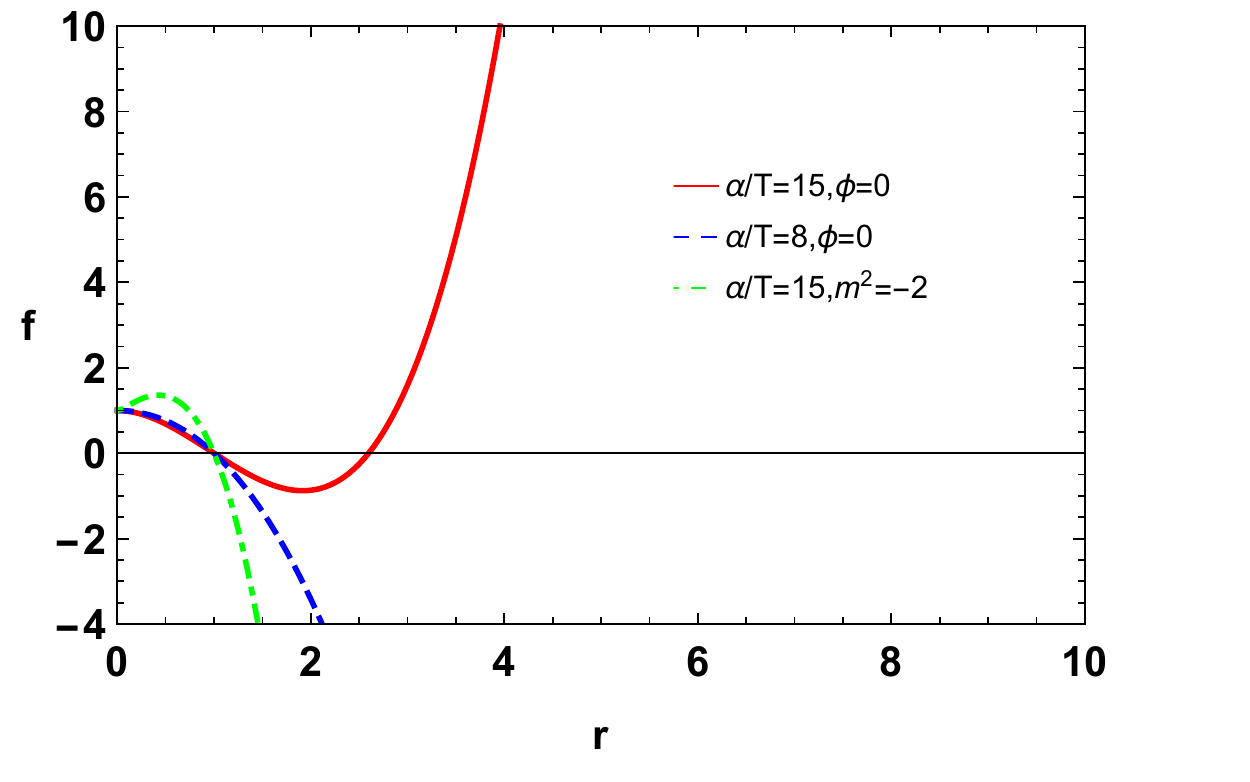}
	\includegraphics[scale=0.6]{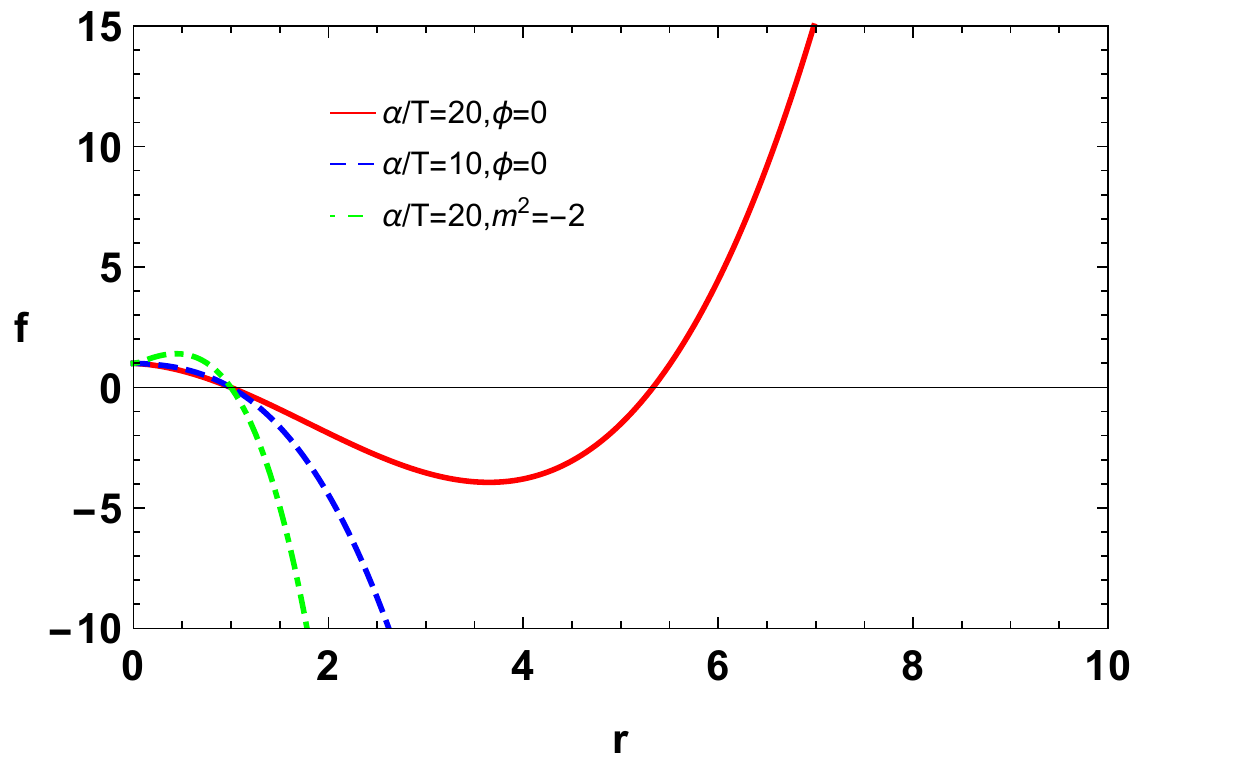}
	\caption{The blackening factor $f(r)$ as a function of $r$ from the boundary $r=0$ to near the black hole singularity $r=\infty$. The \textbf{left panel} refers to the type I linear axion case with $K=X$, while the \textbf{right panel} to the dRGT type II scenario $K=a_{1}\sqrt{X}+a_{2} X$ with $a_{1}=0.1$ and $a_{2}=0.3$.  \label{TS3}}
\end{figure}
The interior of the black hole depends on the choice of the parameter $\alpha$ which determines the size of the graviton mass and the rate of translations breaking in the dual field theory~\cite{Davison:2013jba}. When $0\leq\alpha<\sqrt{2}/r_{+}$, the black hole has no inner Cauchy horizon inside the event horizon $r_+$, and the interior geometry will end at a spacelike singularity. For other cases, there is typically an inner horizon, for which the singularity is timelike. See the left panel of Fig.~\ref{TS3} for an illustration.

For the dRGT case with $K=a_{1}\sqrt{X}+a_{2}X$ (type II), the black brane solution takes the form
\begin{equation}\label{massivebh1}
	f(r)=1-\left(\frac{r}{r_{+}}\right)^3-\frac{a_{1}\,\alpha}{4}\left(r- r_{+}\left(\frac{r}{r_{+}}\right)^3\right)-\frac{a_{2}\,\alpha^2}{2}\left(r^2-r_{+}^2\left(\frac{r}{r_{+}}\right)^3\right)\,.
\end{equation}
In this case, the black hole has both an event horizon and an inner horizon, except when
\begin{equation}
	a_{1}<\,\frac{4}{r_+\,\alpha}\,\quad\text{and}\,\quad a_2\,<\,\frac{4-a_1\,r_+\,\alpha}{2\,r_+^2\,\alpha^2}
\end{equation} 
for which the inner horizon is absent (see the right panel of Fig.~\ref{TS3}).

\subsection{Proof of no inner-horizon}\label{prpr}	%
Following Ref.~\cite{Hartnoll:2020rwq}, let's prove the absence of an inner-horizon in the model defined in Eq.~\eqref{eq6}. Suppose the existence of an inner horizon at $r_{\mathcal{I}}>r_+$, for which one has $f(r_{\mathcal{I}})=f(r_+)=0$. From Eq.~\eqref{eq2}, we obtain 
\begin{equation}\label{eq3}
\int_{r_{+}}^{r_{\mathcal{I}}} \left(\frac{f e^{-\frac{\chi}{2}} \phi \phi'}{r^2}\right)'dr=\int_{r_{+}}^{r_{\mathcal{I}}} e^{-\frac{\chi}{2}}{r^{-4}} \left(\diff{V}{\phi^2} \phi^2+r^2 f (\phi')^2\right)dr\,.
\end{equation}
Clearly, the left hand side yields 
\begin{equation}
\int_{r_{+}}^{r_{\mathcal{I}}} \left(\frac{f e^{-\frac{\chi}{2}}\phi \phi'}{r^2}\right)'dr=\frac{f e^{-\frac{\chi}{2}}\phi \phi'}{r^2}\Big|_{r_{\mathcal{I}}}^{r_{+}}=0\,,
\end{equation}
where we have used the fact that $f(r_{\mathcal{I}})=f(r_{+})=0$. On the other hand, since $f(r)$ is negative between the two horizons, when we take $\diff{V}{\phi^2}< 0$ for every value of $\phi$, the integrand in the right hand is negative. Therefore, the only way there can be two horizons is for $\phi=0$. The presence of the scalar hair, $\phi\neq 0$, necessarily removes the inner horizon. 
For example, for the free scalar case with $V=m^2 \phi^2$, the inner horizon will not appear when $m^2<0$ corresponding to relevant operators in the boundary theory. Notice that the cases of irrelevant operators $m^2>0$ or any other possible potential with $\diff{V}{\phi^2}$ positive evade this proof and therefore the inner horizon could appear again.

\section{Collapse of the Einstein-Rosen bridge}\label{sec:ER}

As we have proved in the last section, the black hole with non-trivial scalar hair has no inner horizon, and the black hole interior ends at a spacelike singularity at $r \to \infty$. Following the spirit of Ref.~\cite{Hartnoll:2020rwq}, we do expect to see a crossover around $r_{\mathcal{I}}$, \emph{i.e.} the position of the would-be inner horizon. In particular, in absence of scalar hair, $\phi=0$, there is typically an inner horizon at $r=r_{\mathcal{I}}$ for the black hole solutions~\eqref{massivebh} and \eqref{massivebh1} in the massive gravity theory considered. However, no matter how small the scalar hair is, it has strong non-linear effect closed to the would-be inner Cauchy horizon of ~\eqref{massivebh} and \eqref{massivebh1}, triggering an instability of the latter.
 
This crossover can be obtained analytically when the scalar field is small. Note however that the spacetime dynamics is highly nonlinear in the small scalar field limit. This fact is associated with a collapse of the Einstein-Rosen bridge between the two asymptotic boundaries\,\footnote{In the black hole interior, $g_{tt}$ is an indicator of the measure for the spatial $t$ coordinate that runs along the wormhole connecting the two exteriors of the black hole, \emph{i.e.} the Einstein-Rosen bridge. A quick decrease in $g_{tt}$ near the would-be inner horizon is thus considered as a collapse of the Einstein-Rosen bridge for a fixed coordinate separation $\Delta t$.}. The main idea of Ref.~\cite{Hartnoll:2020rwq} is that for vanishing small scalar field the instability becomes so fast that one can essentially keep the $r$ coordinate fixed. Let's set $r= r_{\mathcal{I}}+\delta r$, so that $f$, $\chi$ and $\phi$ are now functions of $\delta r$, while any explicit factors of $r$ in the equations of motion (\ref{eq2})- (\ref{eq5}) are set to $r_{\mathcal{I}}$. 

For simplicity, we consider $V(\phi^2)=m^2 \phi^2$. Around the location of the would-be inner horizon, $r=r_{\mathcal{I}}$, the mass of the scalar field can be neglected in (\ref{eq2}) and (\ref{eq5})~\cite{Hartnoll:2020rwq}. By using this approximation, one obtains
\begin{equation}\label{EEq1}
	\left(e^{-\chi/2} f \phi'\right)' = 0, \hspace{0.5cm} 4 r_{\mathcal{I}} f' = 2 r_{\mathcal{I}}^2  f (\phi')^2 + 2 K(\alpha^2 r_{\mathcal{I}}^2)- 12, \hspace{0.5cm} \chi' = r_{\mathcal{I}} (\phi')^2\,.
\end{equation}
Integrating the first equation and writing $\phi' = - c_1 (K(z_{\mathcal{I}}^2 \alpha^2) - 6)^{1/2} e^{\chi/2}/f$ with $c_{1}$ an unspecified constant, one finds the general solution to above equations. In particular, the metric component $g_{tt} = -f e^{-\chi}/r_{\mathcal{I}}^2$ is found to obey~\cite{Hartnoll:2020rwq}
\beq\label{eq:gtt}
c_1^2 \log (g_{tt}) + g_{tt} = -\frac{r_{\mathcal{I}}}{2} \, c_2^2(\delta r + c_3) \,,
\eeq
with $c_2$ and $c_3$ two different constants. Making use of Eq.~\eqref{EEq1} and the above solution, one obtains
\beq\label{eq:phichi}
\phi = - \frac{2 c_1}{r_{\mathcal{I}} c_2} \log \left(c_4 \, g_{tt}\right) \,, \qquad e^{-\chi} = \frac{2 z_{\mathcal{I}}^4}{c_1^2 (K(r_{\mathcal{I}}^2 \alpha^2)-6)} (\phi')^2 g_{tt}^2 \,.
\eeq
Clearly, the scalar field shows a logarithmic growth as the metric component $g_{tt}$ becomes small close to the would-be inner horizon.

\begin{figure}
	\includegraphics[width=0.45\linewidth]{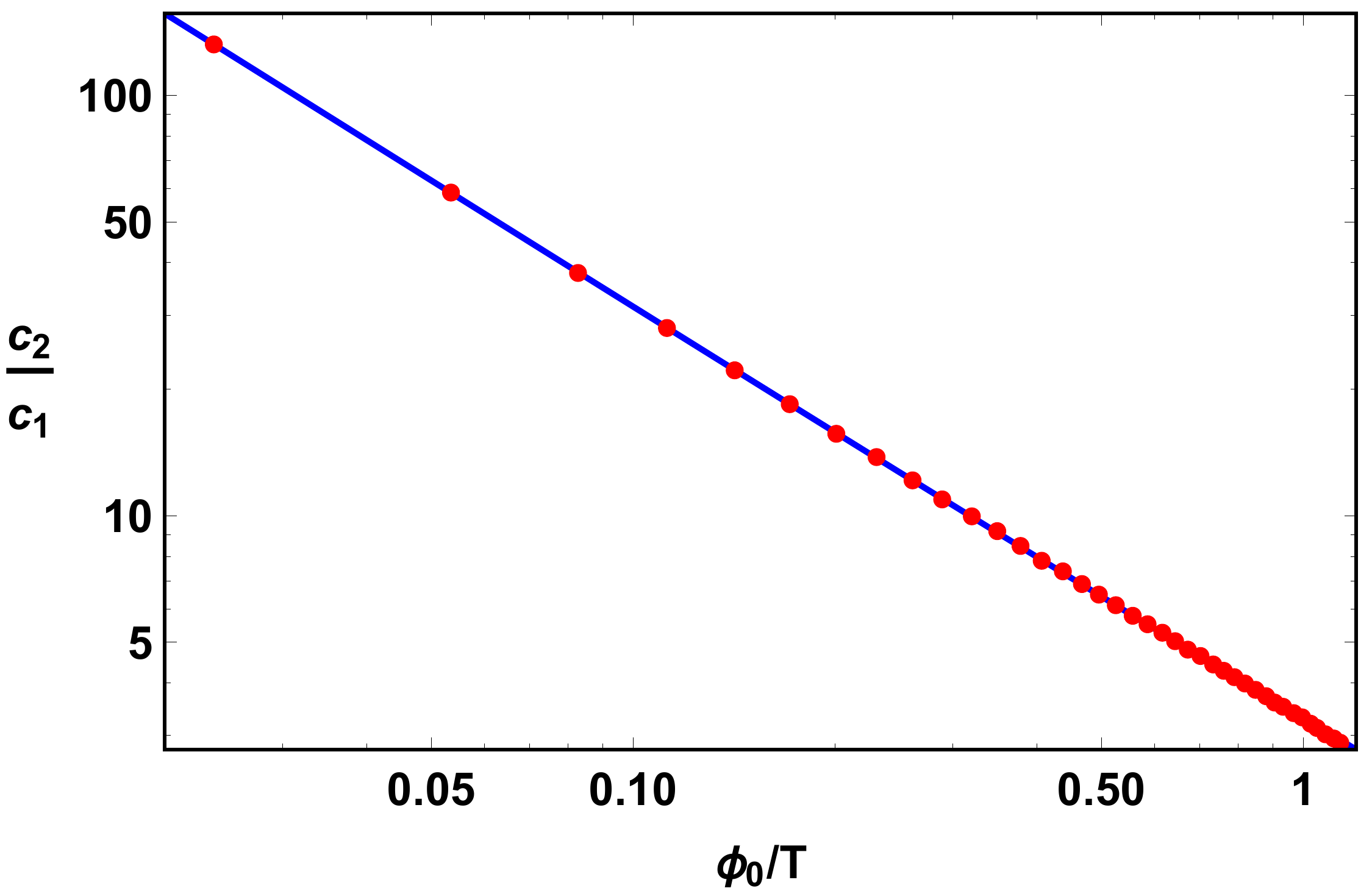} \qquad
	\includegraphics[width=0.45\linewidth]{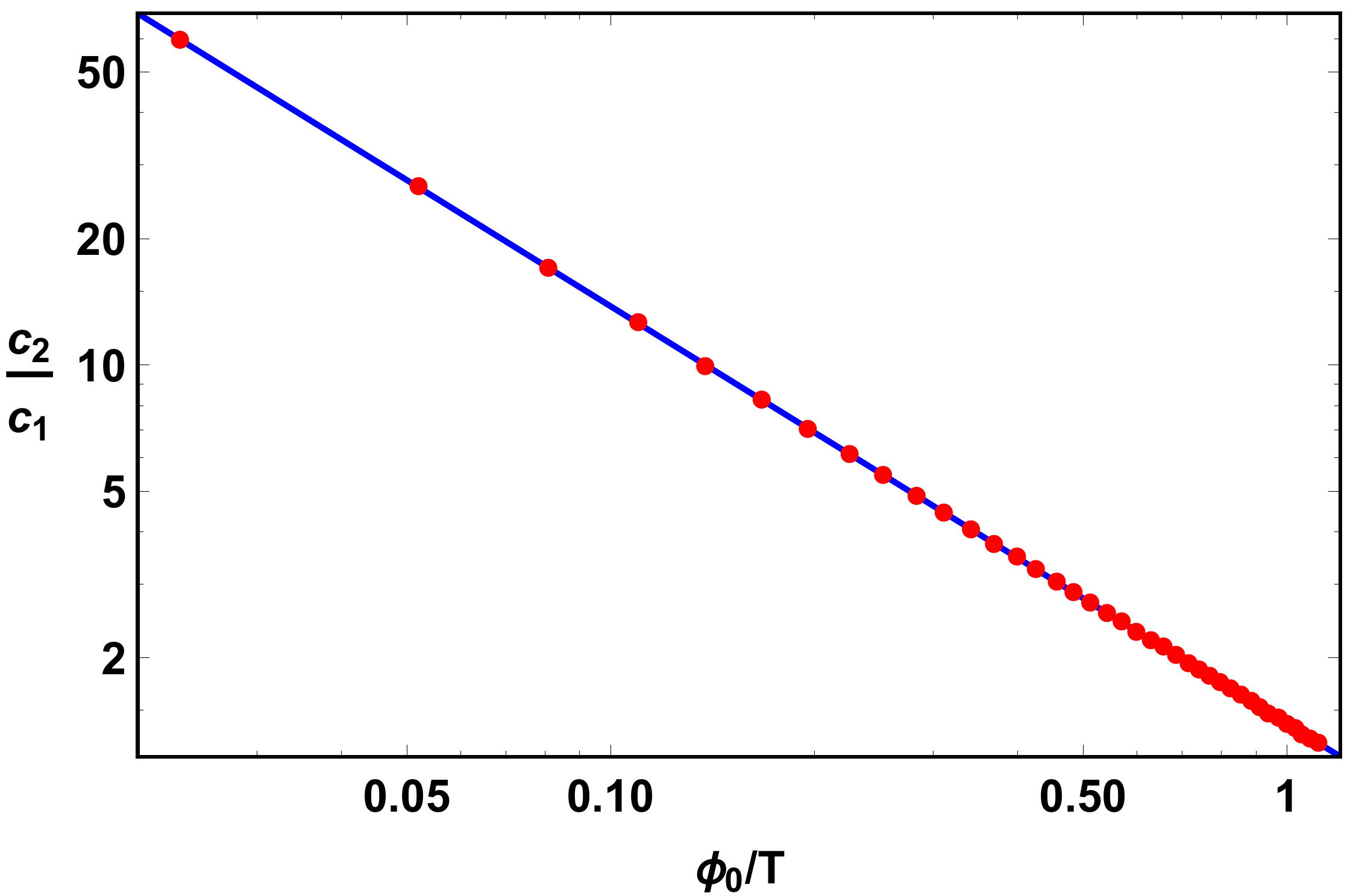}
	\caption{The ratio $c_{2}/c_{1}$ as a function of the dimensionless scalar source $\phi_0/T$. The blue lines show the theoretical prediction $c_{2}/c_{1}=a+b T/\phi_{0}$. $\textbf{Left panel}$: Type I model $K=X$ with $\alpha/T=19.8$. $\textbf{Right panel}$: Type II case with $K=0.1\sqrt{X}+0.3 X$. We have fixed $\alpha/T=24.95$.
		\label{figrphi223}}
\end{figure}

\begin{figure}
	\includegraphics[width=0.46\linewidth]{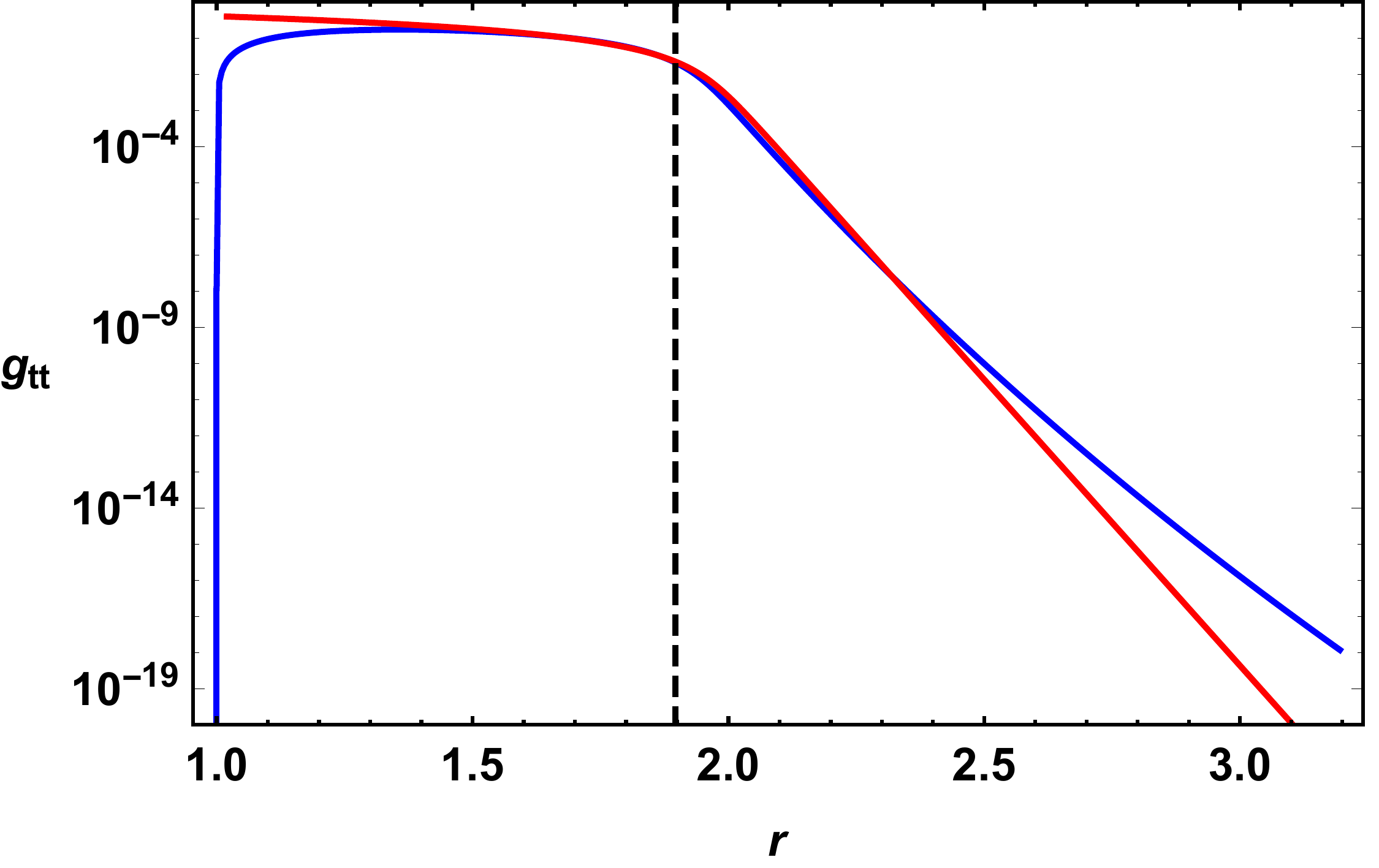}
	\includegraphics[width=0.46\linewidth]{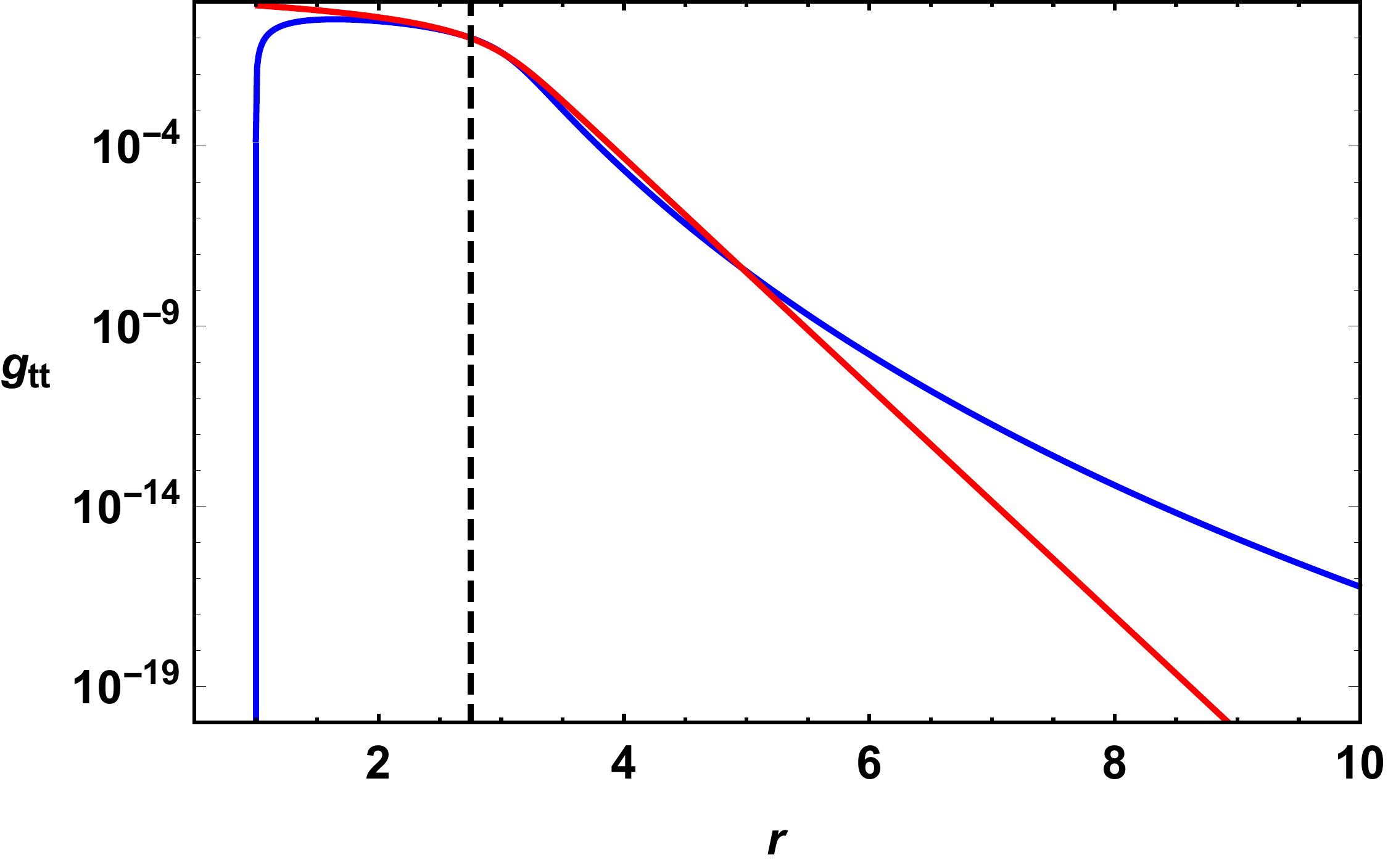}
	\caption{The sudden crossover of the metric component $g_{tt}$ at the would-be Cauchy horizon. The numerical solution of the equations of motion is depicted by the blue curve, while the red line is a fit to the analytic form~\eqref{eq:gtt} valid around $r\sim r_{\mathcal{I}}$.   $\textbf{Left panel}$: The Type I model with $\alpha/T=19.8$. $\textbf{Right panel}$: The Type II case with $K=0.1\sqrt{X}+0.3 X$. We have fixed $\alpha/T=24.95$. The location of the would-be inner horizon $r_{\mathcal{I}}$ is denoted by the dashed vertical line.
		\label{figrphi22}} 
\end{figure}

As pointed out in~\cite{Hartnoll:2020rwq}, the value of $c_2/c_1$ will be large when the boundary source for the scalar $\phi$ is small. We check numerically in Fig.~\ref{figrphi223} that indeed the ratio $c_2/c_1$ scales as $\sim T/\phi_{(0)}$ as the source $\phi_{(0)}/T \to 0$.  Therefore, the term $(c_2/c_1)^2 \delta r$ in (\ref{eq:gtt}) can become very large, which in turn allows the metric component $g_{tt}$ to undergo a suddenly change in the vicinity of $r_{\mathcal{I}}$.
Near the would-be inner horizon, for $r< r_{\mathcal{I}}$ (or $\delta r<0$), $g_{tt} \propto \frac{r_{\mathcal{I}} c_{2}^2}{2} \delta r$ vanishes linearly towards $r_{\mathcal{I}}$, while for $r > r_{\mathcal{I}}$ (or $\delta r>0$) one observes a rapid collapse of $g_{tt}$ to an exponentially small value, \emph{i.e.} $g_{tt} \propto e^{-(c_2/\sqrt{2}c_1)^2 r_{\mathcal{I}} \delta r}$. Note that this collapse occurs over a coordinate range $\Delta r = (c_1/c_2)^2$ as illustrated in \cite{Hartnoll:2020fhc}. This behavior is illustrated in Fig.~\ref{figrphi22} for a small value of the scalar source $\frac{\phi_{0}}{T}=0.5$ in both Type I and II cases. In addition, the scalar field derivative changes from $\phi'=c_{1}/(c_{2} |\delta r|)$ for $r<r_{\mathcal{I}}$ to $\phi'=c_{1}/c_{2}$ for $ r>r_{\mathcal{I}}$, revealing the highly nonlinear nature of this rapid transition.


%
%

\section{Thermal holographic flows and the Kasner singularity}\label{sec:kasner}
After studying the internal structure of our hairy black holes as well as the the collapse of the Einstein-Rosen bridge that occurs at the location of $r_{\mathcal{I}}$ of the would-be inner horizon, we now construct the holographic flows sourced by the scalar field $\phi$ and analyze, in particular, the asymptotic behavior near the spacelike singularity. 

\begin{figure}
	\includegraphics[width=0.42\linewidth]{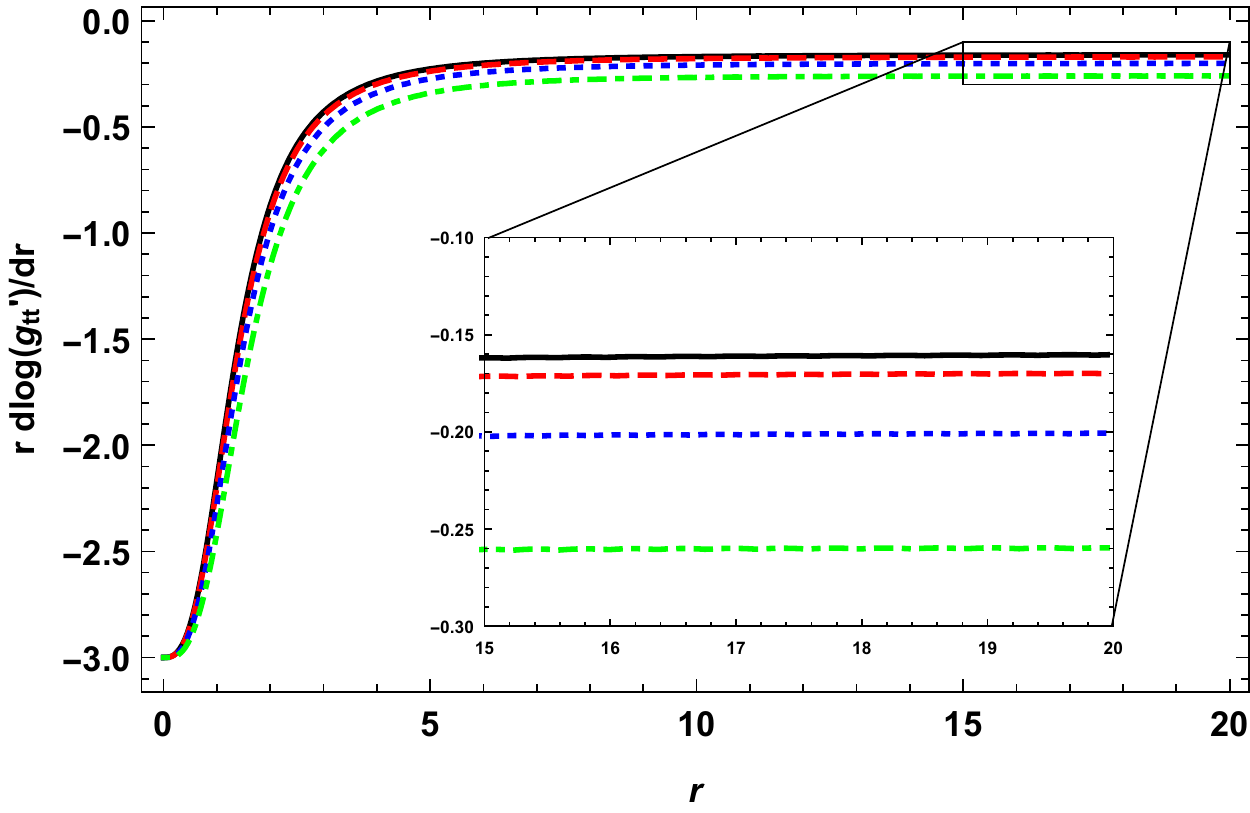}\qquad
	\includegraphics[width=0.42\linewidth]{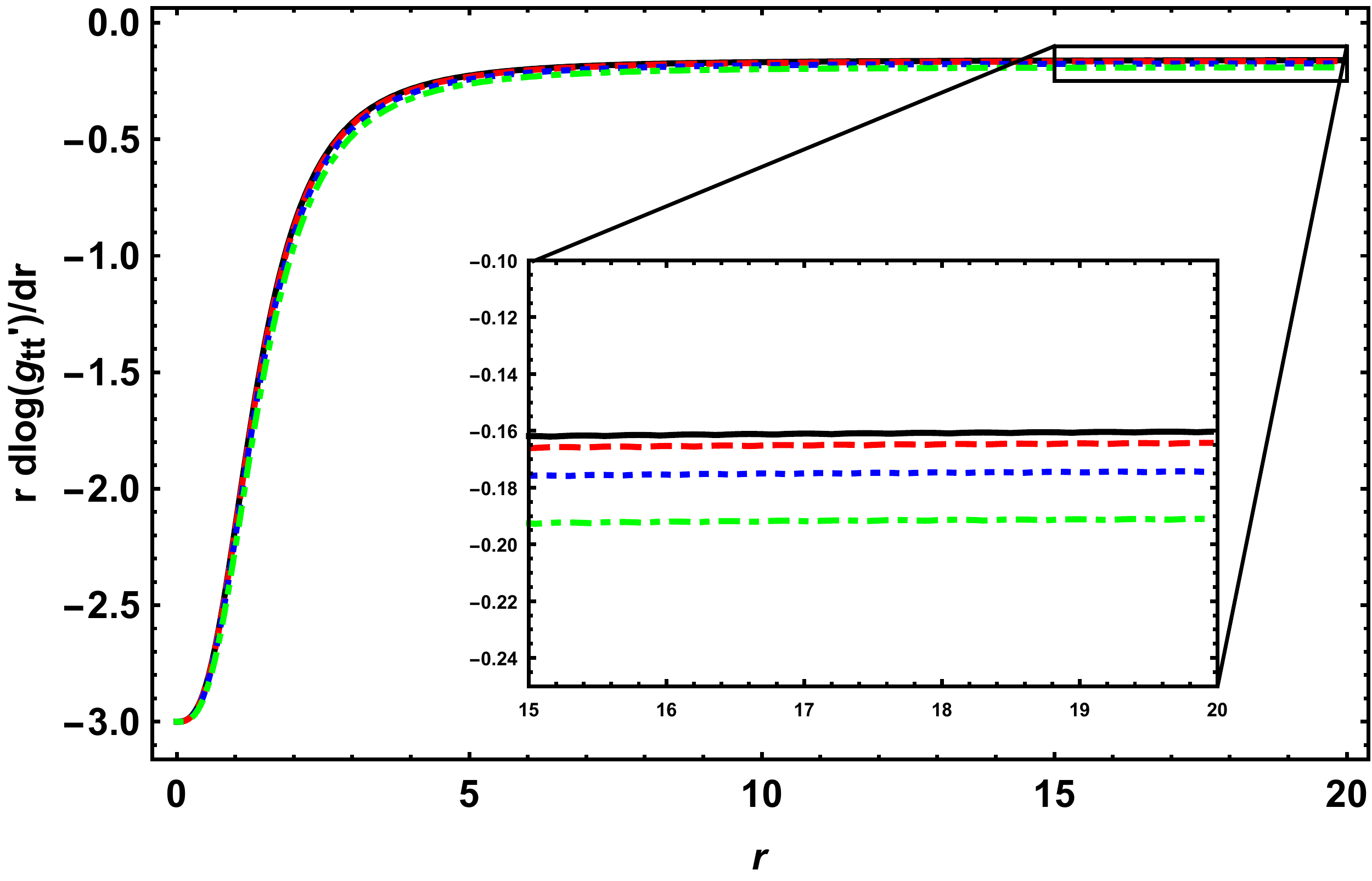}
	
	\vspace{0.2cm}
	
	\includegraphics[width=0.44\linewidth]{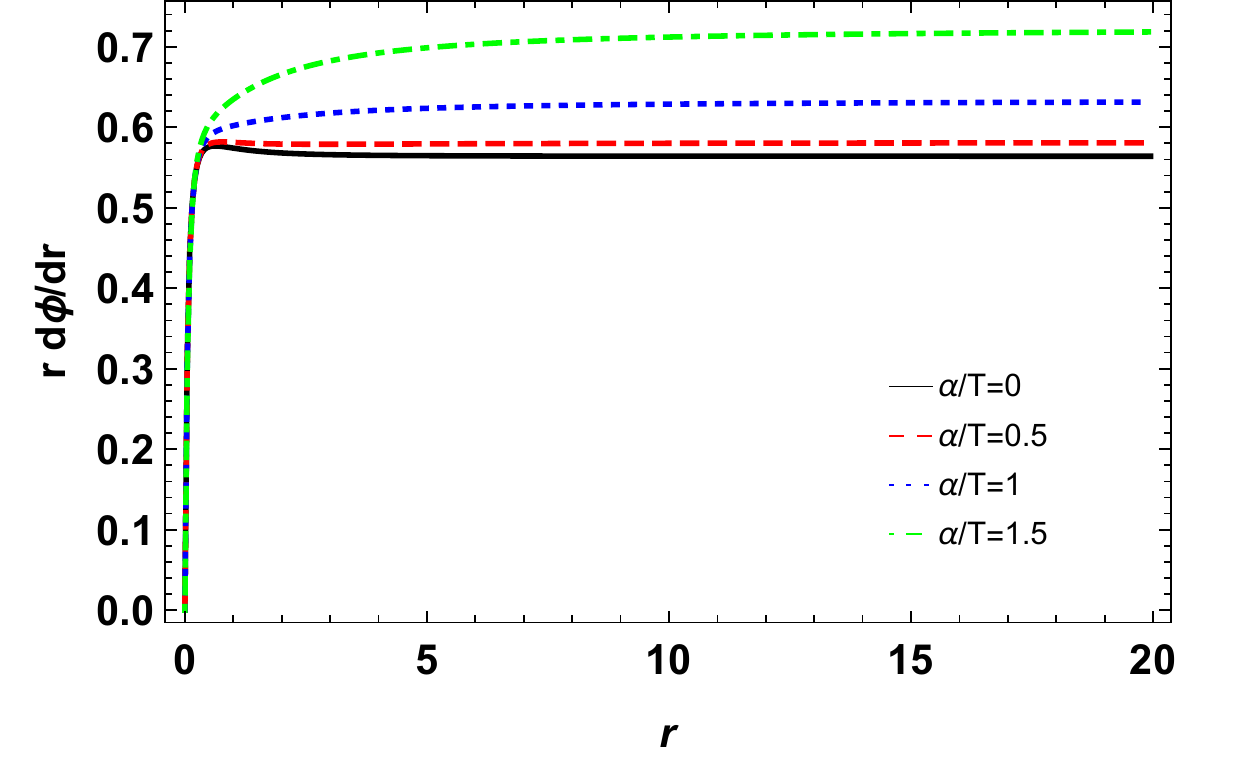}\qquad
	\includegraphics[width=0.44\linewidth]{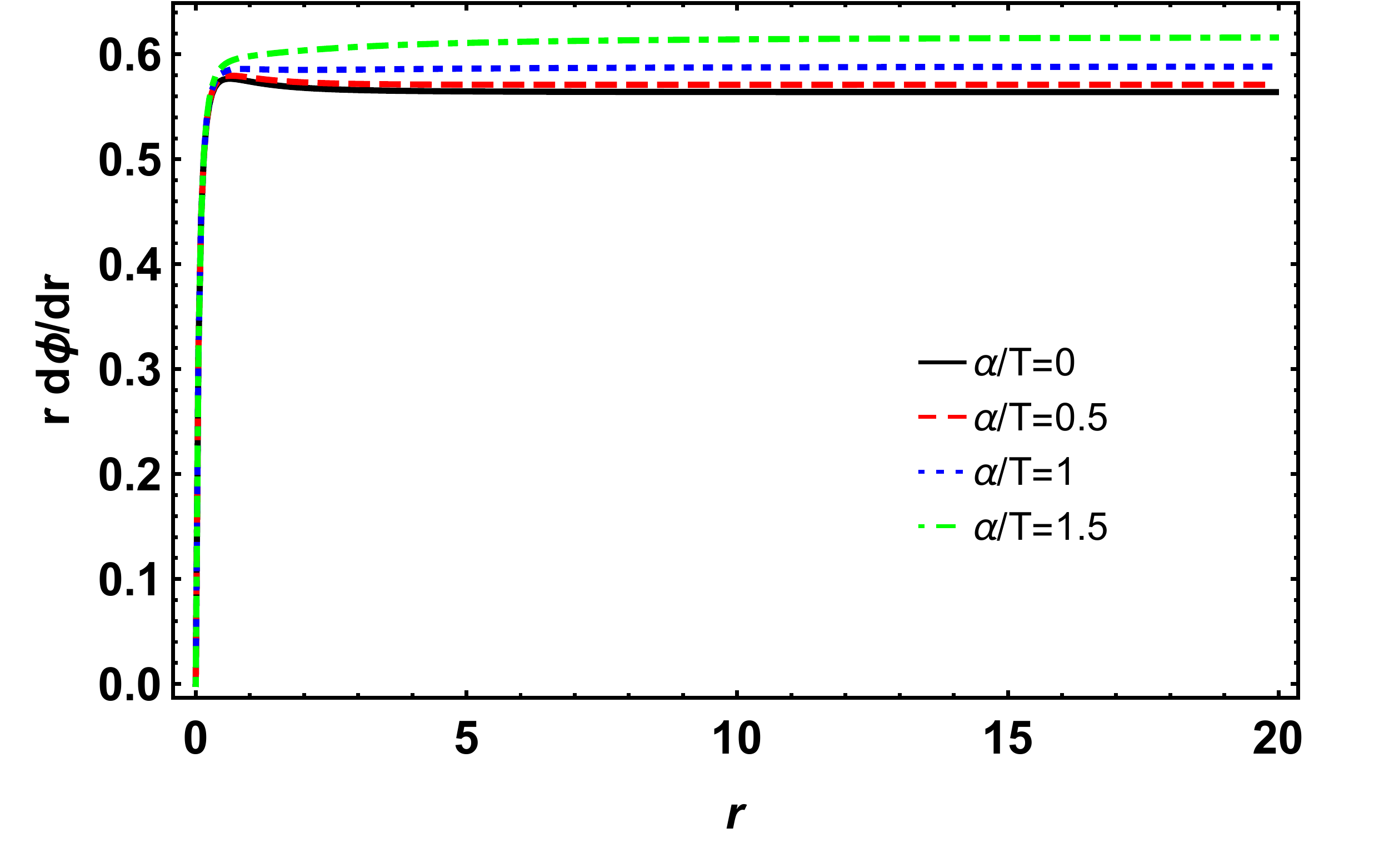}
	
	\vspace{0.2cm}
	
	\includegraphics[width=0.45\linewidth]{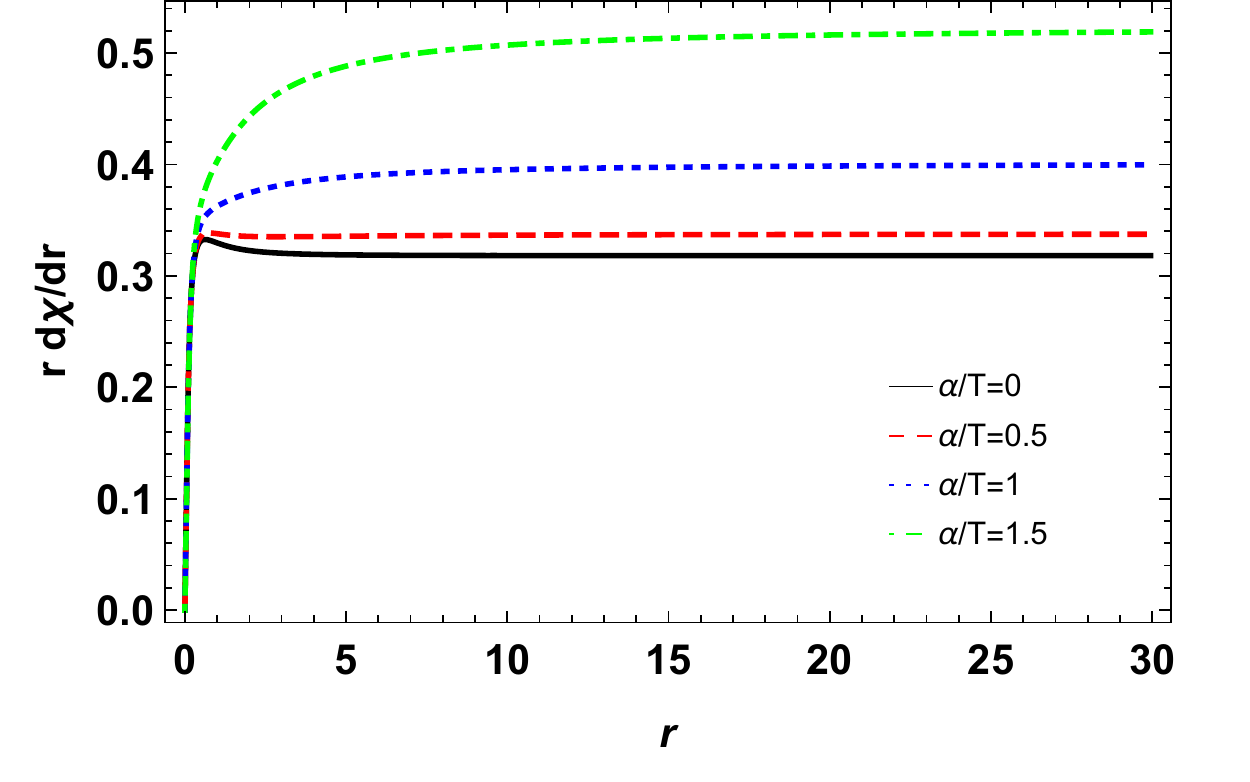}\qquad
	\includegraphics[width=0.45\linewidth]{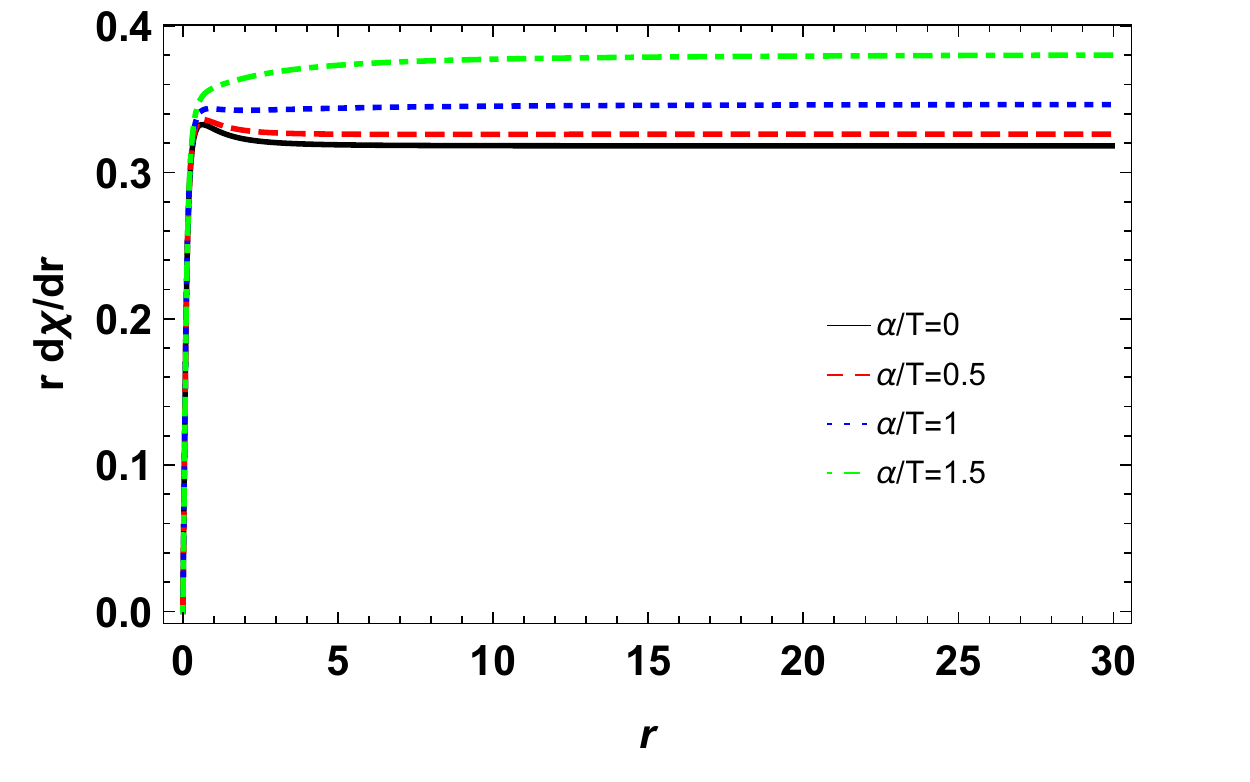}
	\caption{The functions $r \,dX/dr$ with $X=\{\phi,\chi,\log g_{tt}'\}$.
	$\textbf{Left panel}$: Type I case. $\textbf{Right panel}$: Type II case with $a_{1}=0.1$ and $a_{2}=0.3$. 
	The solid dark, dotted blue, dashed red, and dotted dashed green curves correspond to $ \alpha/T=\{0,0.1,0.5,1\}$, respectively. For these cases, the functions $r \,dX/dr$ tend to be a constant at spacelike singularity. These constants will be determined by the Kasner exponents.  \label{fig2}}
\end{figure}

In Fig.~\ref{fig2} we present the behaviors of the various functions approaching the singularity and for different values of $\alpha/T$ by taking $\phi_{0}/T  = 12.5$ and $V(\phi^2)=-2 \phi^2$. The curves therein provide explicit examples of holographic flows that interpolate from a UV radial scaling to a timelike scaling towards a late time singularity behind the black hole event horizon, when the thermal state of the dual CFT is deformed by a relevant scalar operator.

Interestingly, we find that for both type I and type II models, all curves approach to constant values near the singularity as $r\rightarrow \infty$. Actually, we verify numerically (and validate a posteriori) that for our present cases the mass term of the scalar field and the potential $K$ in the equations of motion can be neglected. Assuming that the contributions from the graviton mass (\emph{i.e.} $\alpha\neq 0$) are negligible, the resulting equations can be solved in generality and the solutions at large $r$ take the following form:
\begin{equation}\label{kasner1}
\phi=2c \log r+..., \hspace{0.5cm} \chi= 4c^2 \log r+..., \hspace{0.5cm} f=-f_{1} r^{3+2c^2}+...\,,
\end{equation}
with $c$ and $f_1$ constants. Therefore, one finds that the geometry near the singularity takes a Kasner form
\begin{equation}\label{kasnerm}
ds^{2} \sim -d \tau^2 + \tau^{2 p_{t}} dt^2+ \tau^{2 p_{x}} \left(dx^2+dy^2\right)\,, \hspace{0.5cm} 	\phi\sim - p_{\phi} \log \tau\,,
\end{equation}
where we have traded the radial coordinate $r$ to the proper time  $\tau$ via $d\tau=\frac{dr}{r \sqrt{f}}$. The Kasner exponents in Eq.\eqref{kasnerm} are given by
\begin{equation}
p_{x}=\frac{2}{3+2c^2}, \hspace{0.5cm} p_{t}=\frac{2c^2-1}{3+2c^2}, \hspace{0.5cm}p_{\phi}= \frac{4c}{3+2c^2}\,.
\end{equation} 
One can immediately verify that the above exponents obey
\begin{equation}
p_{t}+2p_{x}=1, \hspace{0.5cm} p_{\phi}^2+p_{t}^2+2 p_{x}^2=1\,.
\end{equation} 

Before proceeding, let's consider the case without scalar hair, $\phi=0$. Notice that for this case the contributions from the graviton mass are not negligible and therefore Eq.~\eqref{kasner1} is not applicable. As evident from Fig.~\ref{TS3}, for $\phi=0$, the hairless black hole can have different internal structures, depending on the value of $\alpha$. Taking the Type I potential as an example, for $0\leq\alpha<\sqrt{2}/r_{+}$, the black hole~\eqref{massivebh} has no inner Cauchy horizon and the interior is similar to the Schwarzschild case with $p_t=-1/3$. For $\sqrt{2}/r_{+}<\alpha<\sqrt{6}/r_{+}$, there is a smooth inner horizon which has $p_t=1$ in Kasner coordinate. The extremal case $T=0$ is obtained at $\alpha=\sqrt{6}/r_{+}$. Another particular case is $\alpha=\sqrt{2}/r_{+}$ for which there is no inner horizon and $p_t=0$.\\

\begin{figure}
	\includegraphics[scale=0.55]{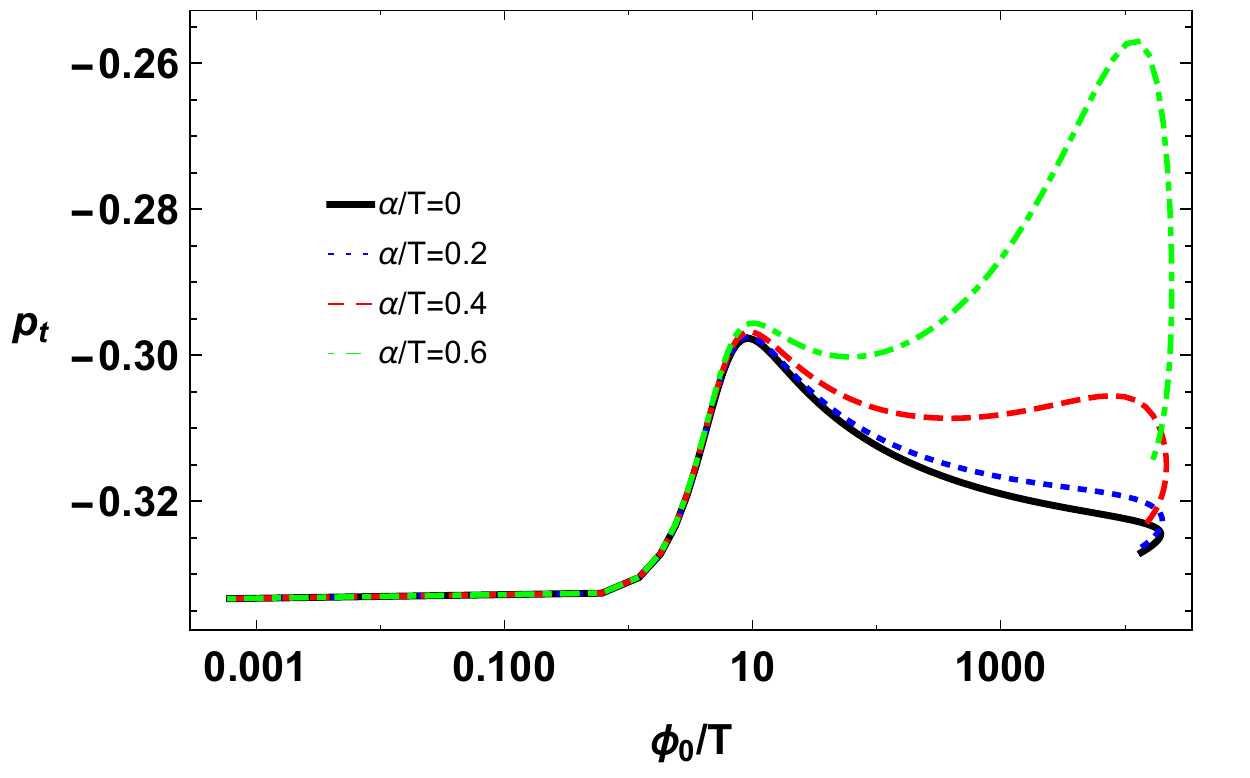}
	\includegraphics[scale=0.55]{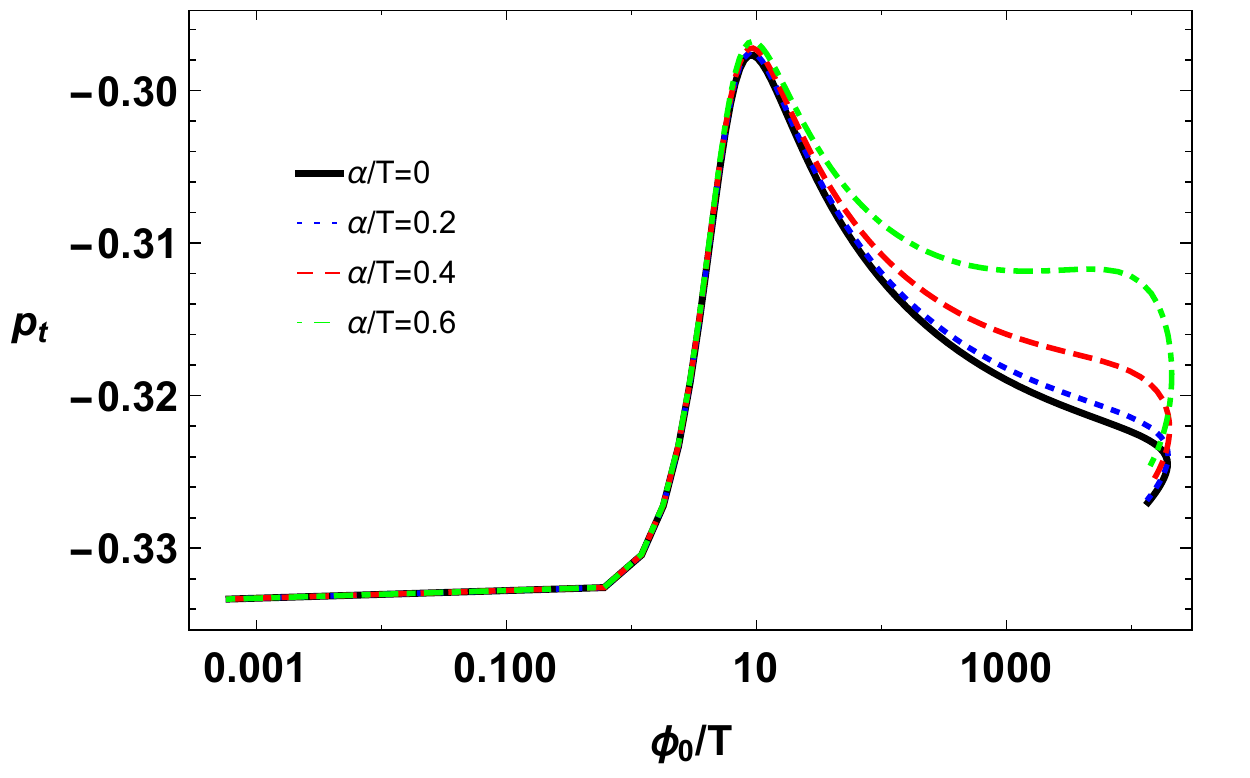}
	\caption{The Kasner exponent $p_t$ of the singularity as a function of the dimensionless scalar source $\phi_{0}/T$ for different values of $\alpha/T$. $\textbf{Left panel}$: Type I case. $\textbf{Right panel}$: Type II case with $a_{1}=0.1$ and $a_{2}=0.3$. As $\phi_{0}/T$ increases for fixed $\alpha/T$, $p_t$ deviates from the value $-1/3$.}\label{figlogpt}
\end{figure}

\begin{figure}
	\includegraphics[width=0.45\linewidth]{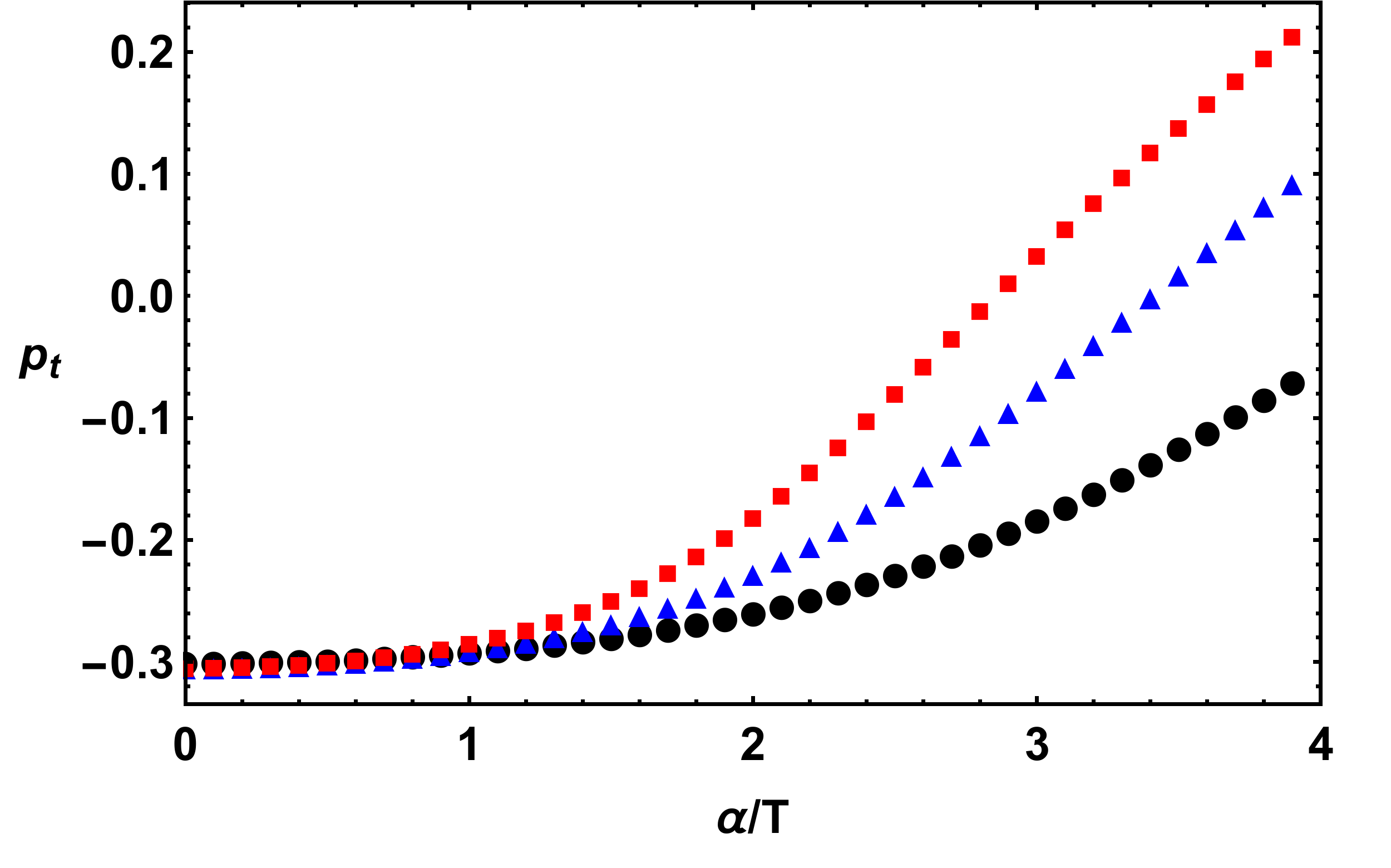}\quad
	\includegraphics[width=0.45\linewidth]{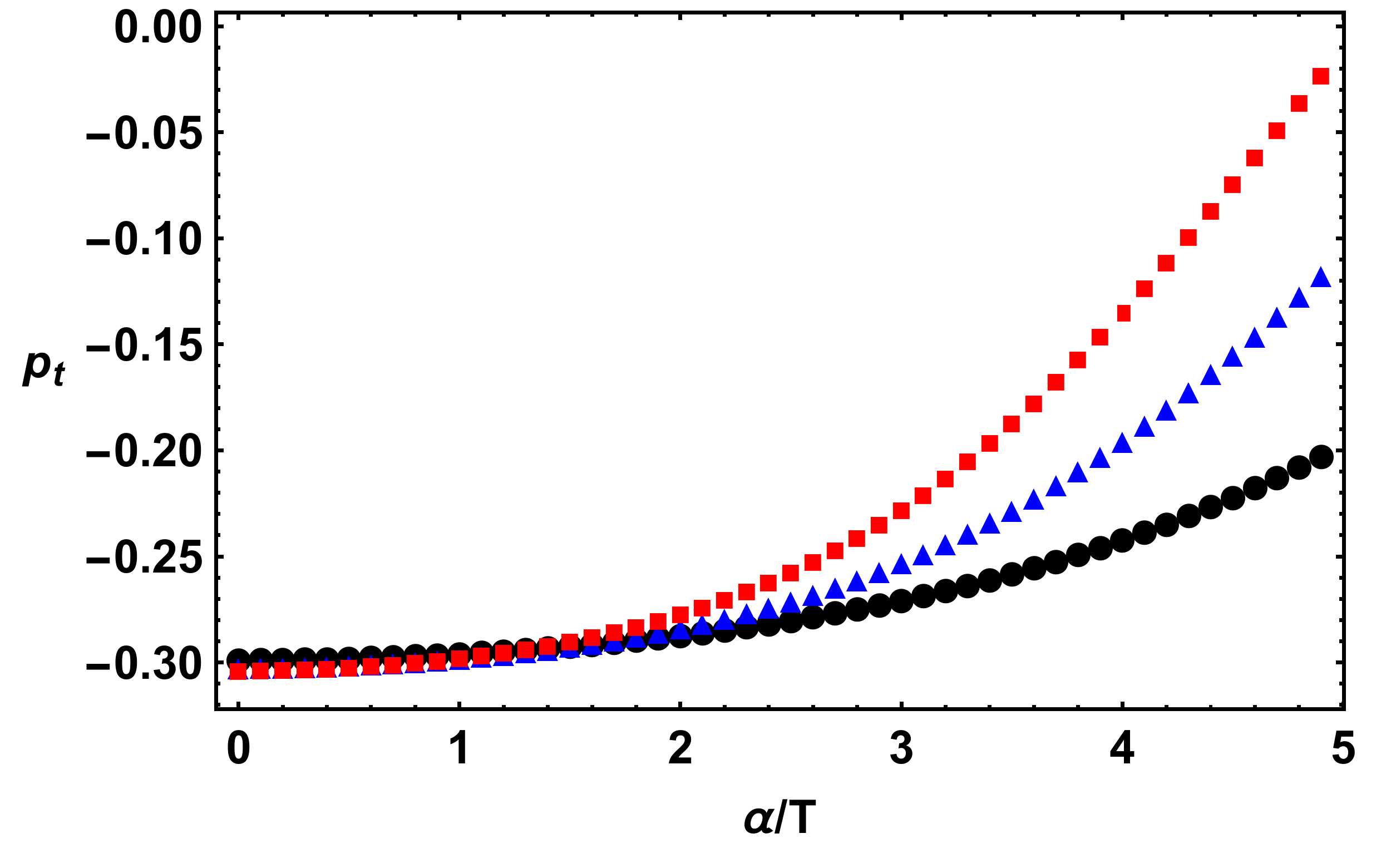}
	\caption{Diagram of the Kasner exponent $p_{t}$ as a function of the parameter $\alpha/T$ for $\phi_{0}/T=\{12.5,15,30\}$ from top to bottom, respectively. Note that the \textbf{left panel} is related to the Type I case with $K=X$, while the \textbf{right panel} is related to the Type II one with $K=a_{1}\sqrt{X}+a_{2} X$ and $a_{1}=0.1$, $a_{2}=0.3$.
	\label{figptalpha}}
\end{figure}

So far, Fig.~\ref{fig2} proves the existence of the already discussed holographic flows from the AdS boundary to an interior Kasner universe. The emergent Kasner scaling is determined by the two  dimensionless CFT parameters $\phi_0/T$ and $\alpha/T$. The Kasner exponent $p_t$ as function of $\phi_0/T$ for different $\alpha/T$ is presented in Fig.~\ref{figlogpt} from which one can see that $p_t$ deviates from $-1/3$ as the source is increased. Therefore, a deformation triggered by the scalar operator changes the near-singularity scaling exponents, signifying a dynamical instability of the singularity for the hairless black hole at later interior time.
The Kasner exponent $p_t$ with respect to $\alpha/T$ with $\phi_0/T$ fixed is shown in Fig.~\ref{figptalpha}. One finds that $p_t$ increases monotonically as $\alpha/T$ is increased. Our numerical data can be fitted well by a quadratic form $c_1+ c_2(\alpha/T)^2$ with $c_1$ and $c_2$ constants.

In the above discussion, we have required 
the kinetic term of scalar field to be dominant with respect to the potential terms $V$ and $K$ of~\eqref{eq6}, or more precisely that 
\begin{equation}\label{potentialcon}
\lim_{r \to \infty} \frac{|V(\phi^2)|}{r^{3+2c^2}}\ll 1 \hspace{0.25cm} \text{and} \hspace{0.25cm} 	\lim_{r \to \infty} \frac{|K(\alpha^2 r^2)|}{r^{3+2c^2}}\ll 1\,.
\end{equation}
In particular, the first condition allows the scalar potential to be an arbitrary polynomial functions of $\phi$. However, as pointed out in~\cite{Cai:2020wrp,An:2021plu}, if one considers a case that diverges exponentially or even worse, the above condition can be violated and the Kasner form would break down. 

Before ending this section, we discuss indeed a scenario in which the Kasner form~\eqref{kasner1} can be violated near the singularity. For illustration, we consider the following form of potentials:
\begin{equation}\label{potential1}
V(\phi^2)=-\phi^2-\cosh(\gamma \phi^2)+1,\quad K(X)=X^n\,,
\end{equation}
with $\gamma>0$ such that $\diff{V}{\phi^2}<0$ to remove any inner horizon, and $n>0$ to avoid the ghost and gradient instabilities~\cite{Baggioli:2014roa}.
Following~\cite{Cai:2020wrp}, we assume we are in the Kasner regime where one has $\phi=2c \ln r$ at large $r$. For the potentials of~\eqref{potential1}, we have 
\begin{eqnarray}
\frac{|V(\phi^2)|}{r^{3+2c^2}}&\sim& \frac{e^{4c^2 \gamma (\ln r)^2}}{r^{3+2c^2}}> \frac{e^{4c^2 \kappa \gamma \ln r}}{r^{3+2c^2}}=r^{4 c^2 \kappa \gamma-3-2c^2}\,,\\
\frac{|K(\alpha^2 r^2)|}{r^{3+2c^2}}&\sim& \frac{(\alpha^2 r^2)^n}{r^{3+2 c^2}}=\alpha^{2 n} r^{2n-3-2c^2}\,,\label{cons1}
\end{eqnarray} 
where $\kappa$ is a constant for which one only requires $\kappa<\ln r$. It is obvious that when $\kappa>\frac{3+2c^2}{4 c^2 \gamma}$ the the first constraint in Eq.~\eqref{potentialcon} is no longer obeyed. In addition, when $n>\frac{3+2c^2}{2}$, the second constraint can be violated as well.

\begin{figure}
	\includegraphics[scale=0.36]{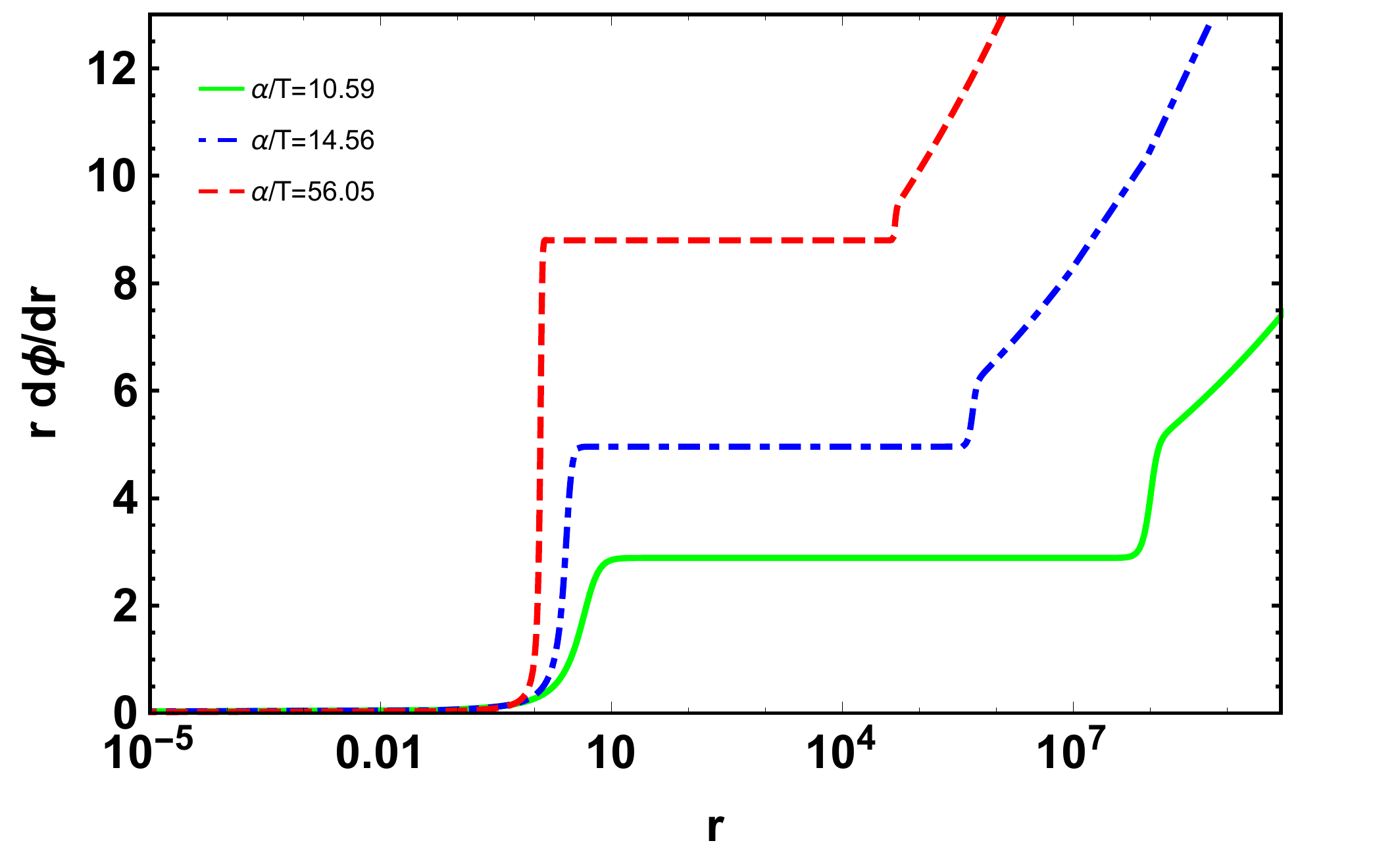}
	\includegraphics[scale=0.36]{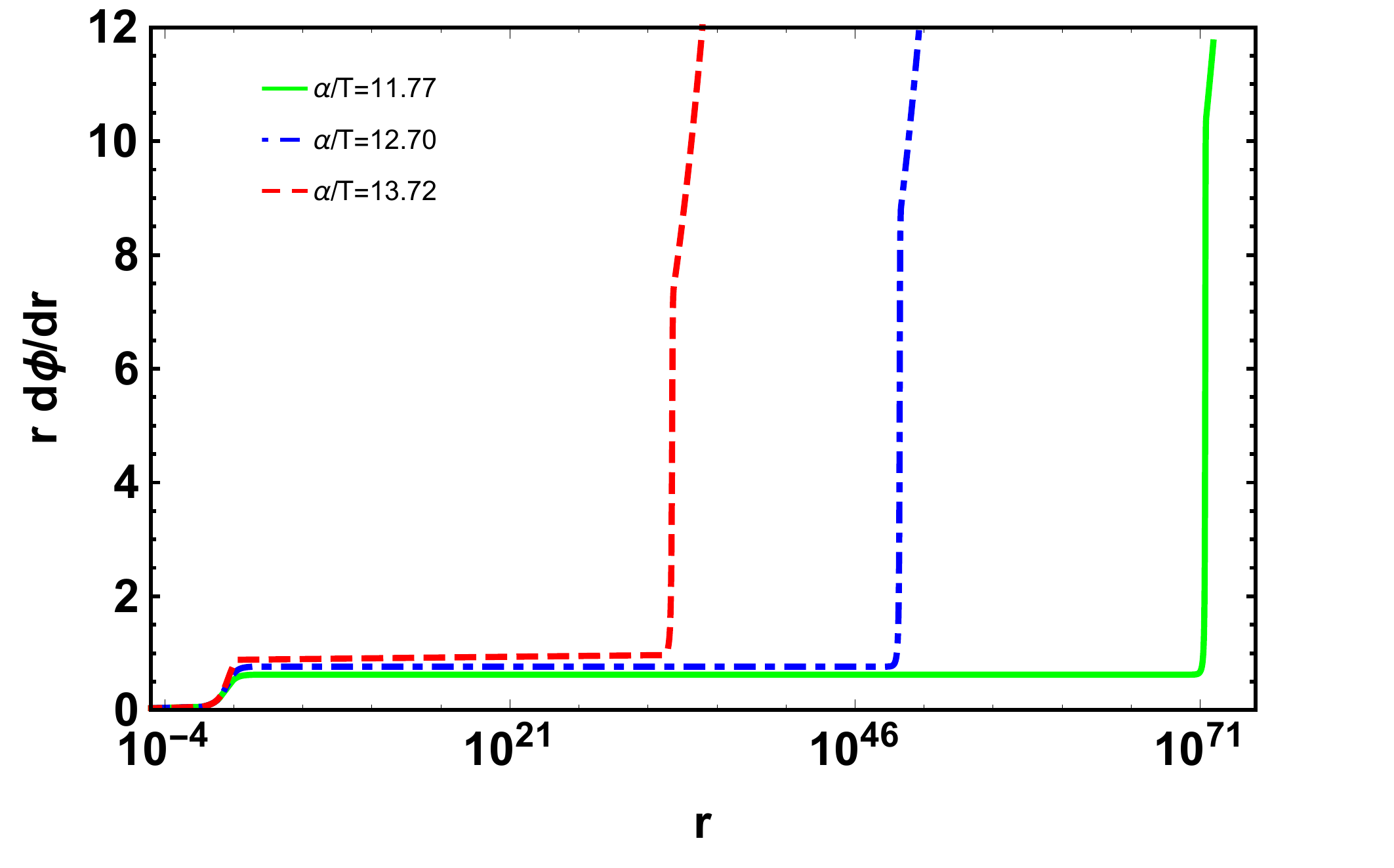}
	\includegraphics[scale=0.58]{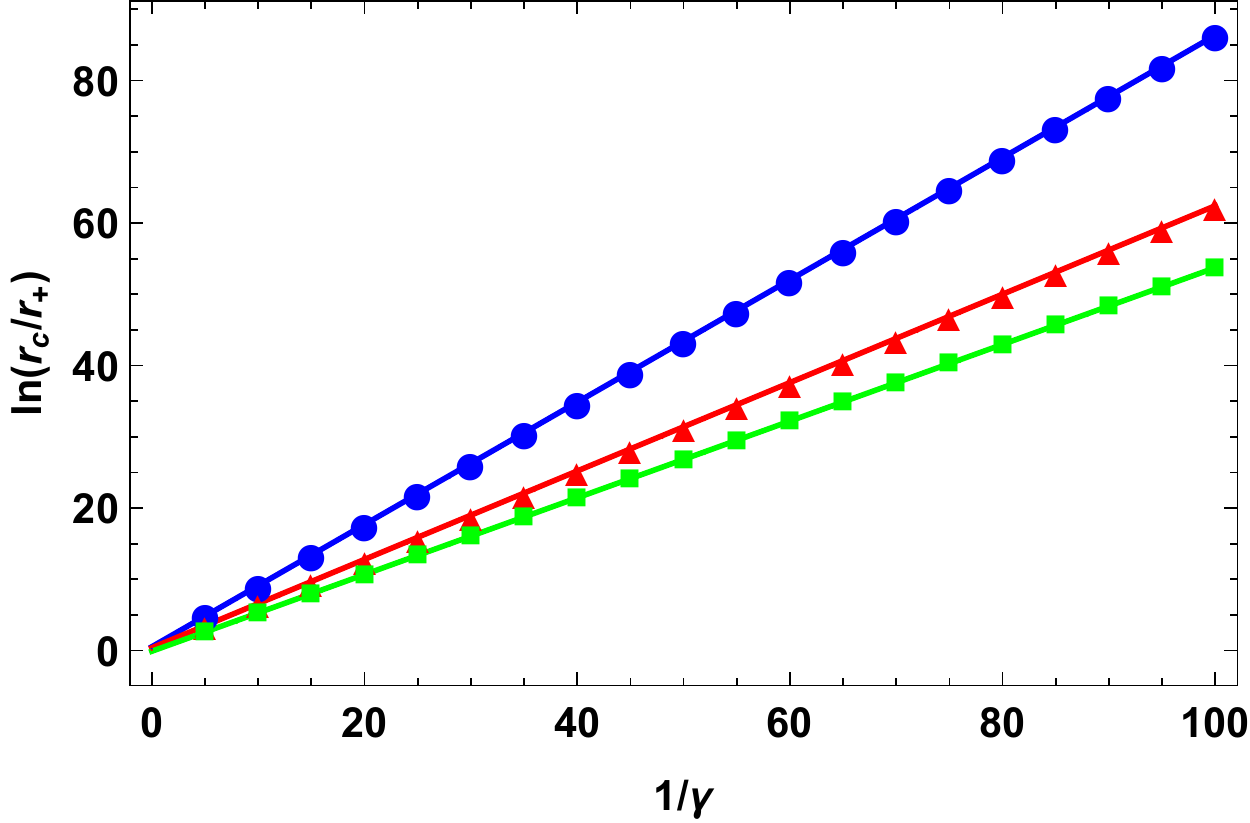}
	\includegraphics[scale=0.58]{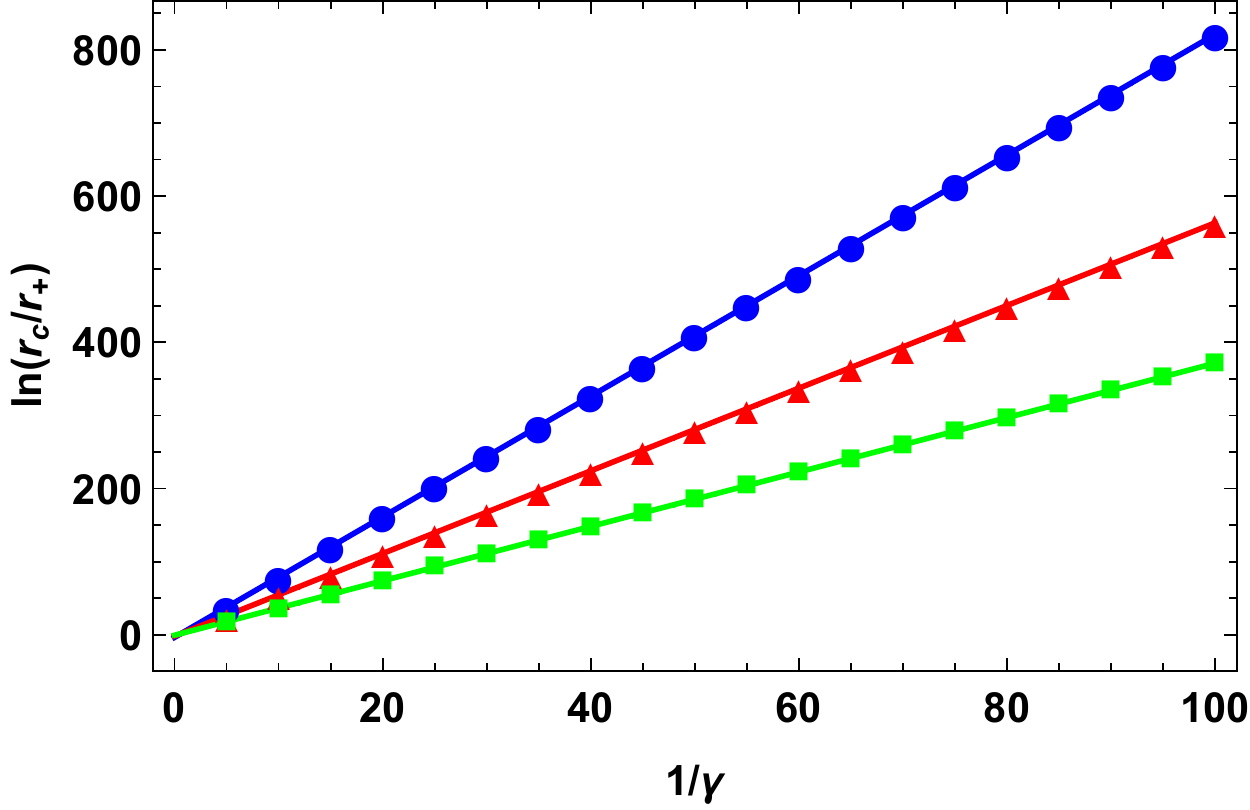}
	\caption{\textbf{Top panel: } The behavior of $r d\phi(r)/dr$ near the singularity for $V(\phi^2)=-\phi^{2}-\cosh(0.05\phi^2)+1$. \textbf{Bottom panel: } The relationship between $r_{c}$ and $\gamma$. Here, we select $\phi(r_{h})=0.5$ and $\chi(r_{h})=0$ with $r_{h}=1$ in both cases. Note that the \textbf{left panels} are related to the case Type I with $K=X$, while the \textbf{right panels} are related to the case Type II with $K=a_{1}\sqrt{X}+a_{2} X$ when one takes $a_{1}=0.1$ and $a_{2}=0.3$. The solid lines are the theoretical predictions presented in Eq.~\eqref{scalingV}.
		\label{figrphi2}}
\end{figure}

\begin{figure}
	\includegraphics[scale=0.3]{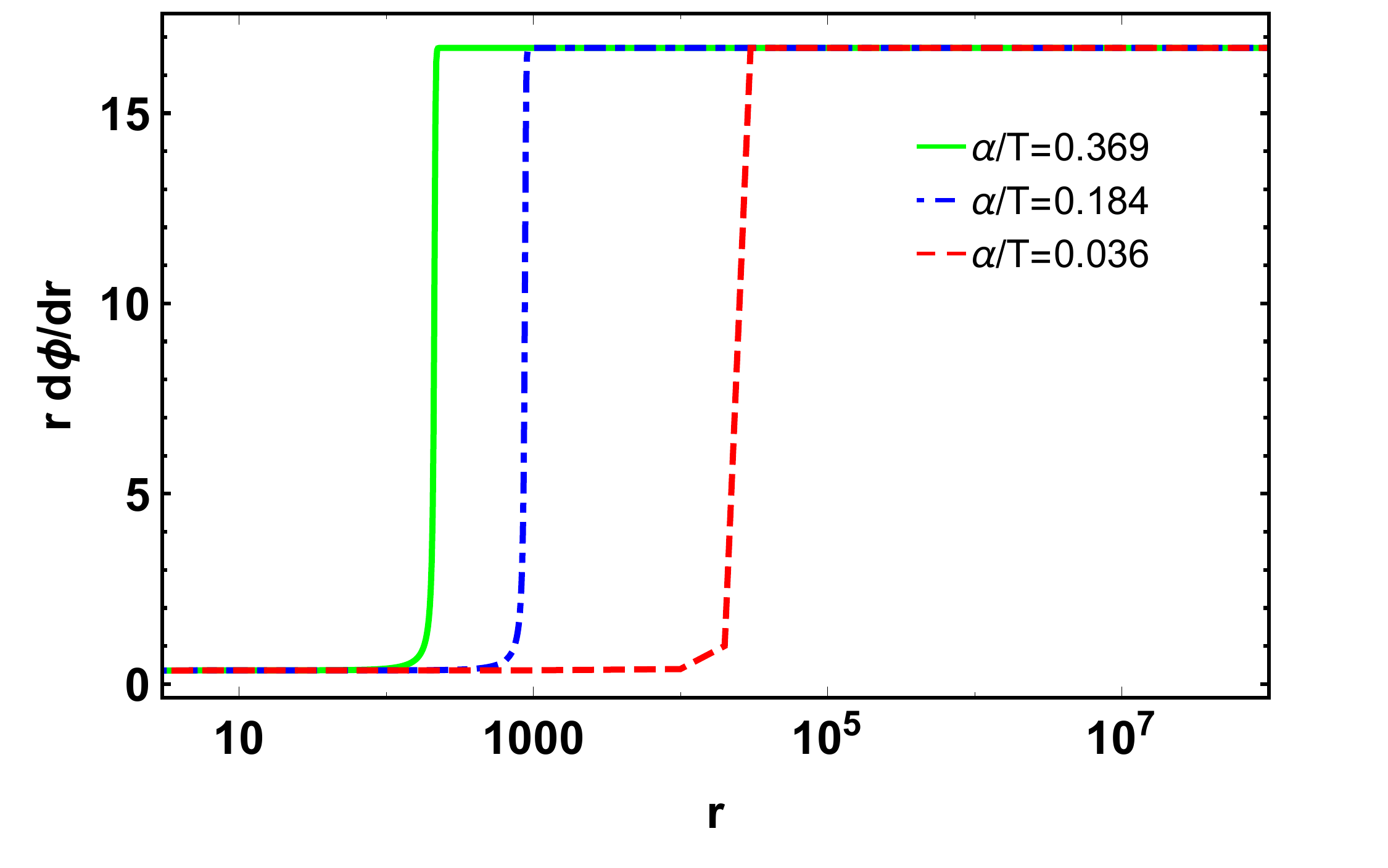}
	\includegraphics[scale=0.5]{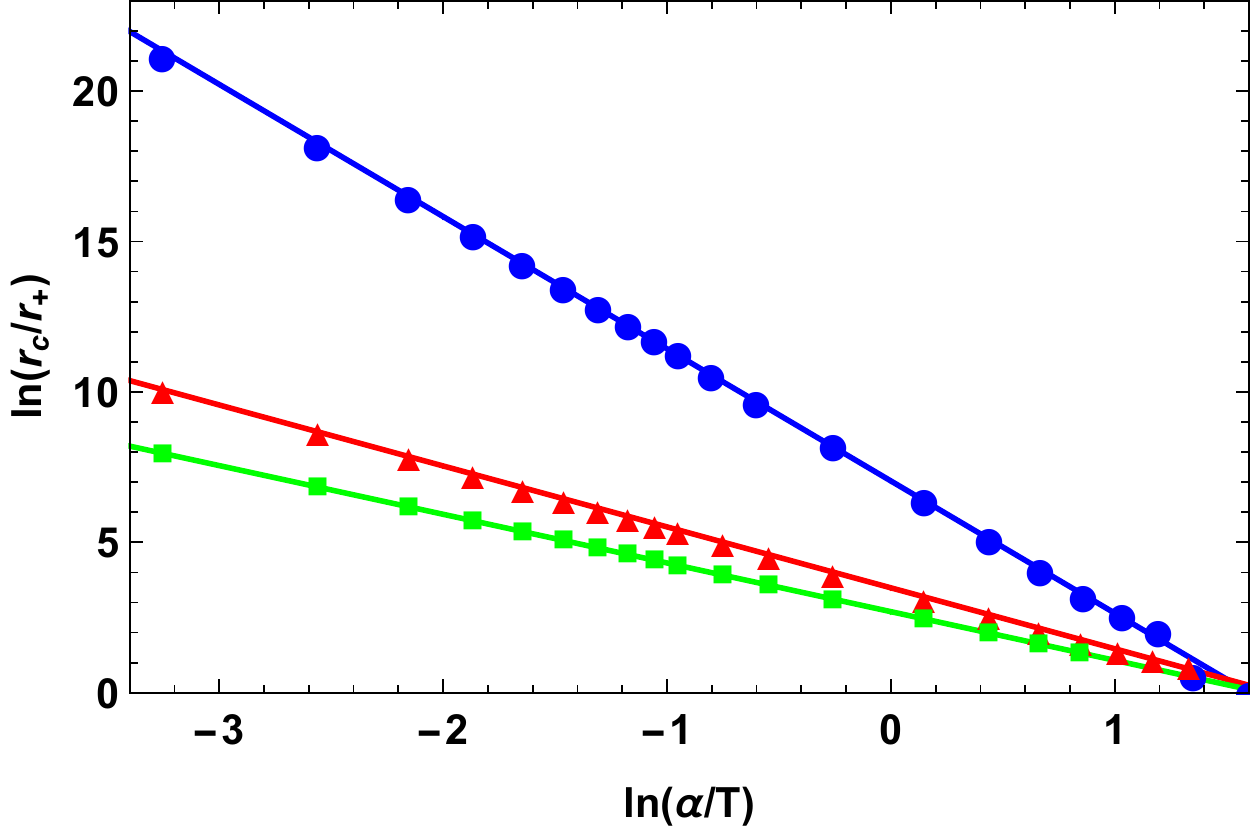}
	\caption{\textbf{Left: }The behavior of $r d\phi/dr$ near the singularity for $V(\phi^2)=-2\phi^{2}$ and $K(X)=X^3$. \textbf{Right: }The relation between the critical point $r_{c}$ and $\ln \alpha$ for different types of massive gravity models $K(X)=\{X^2,X^3,X^4\}$ from top to bottom, respectively. In all cases, one finds numerically a value of the Kasner constant $c \approx 0.17 $ .  Here, we select $\phi(r_{h})=0.5$ and $\chi(r_{h})=0$ with $r_{h}=1$ in both cases. The solid lines show the analytical results in Eq.~\eqref{rca}.
		\label{figrphiX3}}
\end{figure}

The above analysis predicts that, for the scalar potential $V$ of Eq.~\eqref{potential1}, the Kasner form should be violated no matter how small the value of $\gamma$ is. A deviation from the Kasner form is expected beyond a critical point $z_c$ given by
\begin{equation}\label{scalingV}
r_c\sim e^{\frac{3+2c^2}{4\gamma c^2}}\Rightarrow \ln(r/r_+)=\frac{3+2c^2}{4c^2\gamma}+b_\gamma\,,
\end{equation}
with $b_\gamma$ a constant. It is numerically challenging to verify the scaling law~\eqref{scalingV} for small $\gamma$, because one has to solve the equations of motion to sufficiently large $r$. Some examples are shown in Fig.~\ref{figrphi2} from which one observes a noticeable deviation from the asymptotic solution~\eqref{kasner1} towards the singularity. The relationship between $r_c$ at which the Kasner form is violated and $\gamma$ from our numerics is presented in the bottom of Fig.~\ref{figrphi2}. We indeed verify the expected scaling behavior~\eqref{scalingV} for small $\gamma$.

We now turn to the role of $K(X)$. Note that for both Type I and Type II, the second constraint of Eq.~\eqref{potentialcon} is always satisfied. In order to see the deviation from the Kasner form, one should consider other forms of $K(X)$.
The analysis of~\eqref{cons1} suggests to consider $K(X)=X^n$ with $n>\frac{3+2c^2}{2}$. Consequently, we find a deviation from the Kasner behavior~\eqref{kasner1} beyond a position $r_c$ that obeys
\begin{equation}\label{rca}
 \ln \left(\frac{r_{c}}{r_{+}}\right)=-\frac{2n}{2n-(3+2c^2)} \ln \left(\frac{\alpha}{T}\right)+b_{\alpha}
\end{equation}
with $b_{\alpha}$ a constant that depends on the model we take. The case with $K(X)=X^3$ is presented in the left panel of Fig.~\ref{figrphiX3} where we numerically show the behavior of $r d\phi/dr$ at the large $r$. As one can see, there exists a critical value $r_{c}$, beyond which
the the Kasner behavior will be modified. In the right panel of Fig.~\ref{figrphiX3}, we compare the numerical data for $r_c$ versus $\alpha$ with our theoretical prediction~\eqref{rca}. After fitting the coefficient $b_\alpha$, we find that the numerical results agree with the theoretical prediction quite well for all the cases considered, $K(X)=\{X^2, X^3, X^4\}$.

\section{Absence of Josephson oscillations}\label{newsec}
{\color{black}After studying in detail the geometry and the gravitational dynamics inside the BH horizon in our model, it is time to compare our results with those obtained in Ref.~\cite{Hartnoll:2020fhc}. In their case, before reaching the singularity, strong Josephson oscillations in the condensate are observed. As evident from the previous Sections, this is not the case in our setup; let us explain why. In the previous analysis, we ignored the mass of the scalar field in~\eqref{EEq1}. If we keep the mass term, then the first expression in Eq.~\eqref{EEq1} takes the following form:
\begin{equation}\label{myphi}
\frac{e^{-\frac{\chi}{2}}f}{r^2} \Big(\frac{e^{-\frac{\chi}{2}} f \phi'}{r^2}\Big)'=\frac{|m^2 g_{tt}|}{r^4} \phi \,.
\end{equation}
As we have shown, the collapse of the ER brdige occurs in an extremely small range of the $r$ coordinate. Therefore, it is consistent to set $r= r_\mathcal{I}$ in Eq.~\eqref{myphi}. Using this approximation, we obtain the general solution for the scalar field to be
\begin{equation}
\phi=c_1 e^{\beta\int_{r_\mathcal{I}}^r e^{\chi/2}/f dr}+c_2 e^{-\beta\int_{r_\mathcal{I}}^r e^{\chi/2}/f dr}
\end{equation}
where $\beta=\sqrt{|m^2 g_{tt}(r_\mathcal{I})|}$ and $c_1$ and $c_2$ are integration constants. For the charged scalar case considered in Ref.~\cite{Hartnoll:2020fhc} and using the same coordinates system, the analogous equation of motion reads
\begin{equation}\label{chargephi}
\frac{e^{-\frac{\chi}{2}}f}{r^2} \Big(\frac{e^{-\frac{\chi}{2}} f \phi'}{r^2}\Big)'=-\frac{q^2\Phi^2}{r^4} \phi 
\end{equation}
where $q$ and $\Phi$ are respectively the charge of scalar field and the gauge potential. In contrast to~\eqref{myphi}, in which the coefficient of the right hand side is always positive, the one in Eq.~\eqref{chargephi} is negative. This is a fundamental difference which will lead to several consequences. In particular, we can solve Eq.~\eqref{chargephi} and find the oscillating solution discussed in \cite{Hartnoll:2020fhc}
\begin{equation}
\phi=\phi_0\cos\left(|q\Phi(r_\mathcal{I})|\int_{r_\mathcal{I}}^r \frac{e^{\chi/2}}{f}dr+\varphi_0\right)\,,
\end{equation}
where $\phi_0$ and $\varphi_0$ are two constants. As the Reader can notice, the difference sign in the r.h.s. of Eq.~\eqref{myphi} and Eq.~\eqref{chargephi} has the important effect of modifying the oscillating solution $\sim e^{\pm i  k_r r}$ into an exponential one $\sim e^{\pm   k_r r}$ and therefore makes the Josephson oscillations to disappear.\\

Notice how the sign of the mass squared $m^2$ in our setup is fixed by the requirement of having no inner horizon as discussed in Section \ref{prpr}, which concludes our proof for the absence of Josephson oscillations in our case. For completeness, we have verified this fact numerically and we never observed oscillations as expected.\\

It would be interesting to extend our model by charging the bulk scalar field $\phi$ under a U(1) symmetry. In this case, we do expect a competition between the graviton mass and the effective mass coming from the gauge potential in the superconducting phase. Therefore, at least in some regimes, we do expect the Josephson oscillations to re-appear. In the massive gravity theories considered, holographic superconductor models have been already studied in \cite{Baggioli:2015zoa,Baggioli:2015dwa} and they can be directly exploited to answer this question in a more general framework. We leave this new analysis for the future.}

\section{Probes of the black hole interior and Kasner geometry}\label{sec:probe}
In the previous sections, we have shown a rich dynamics in the way the geometry of the BH interior can reach the singularity. For general potentials satisfying the constraint~\eqref{potentialcon}, we have found that the deformation of the thermal CFT by a relevant scalar operator leads to a general Kasner universe at late interior times. As these geometrical features should leave strong imprints to all the observable of the dual CFT probing such a region, in this section, we will use several field theory observables to probe the interior of the hairy black holes in the bulk and our previous findings.

The first probe considered is the entanglement entropy.  The holographic formula for the entanglement entropy of the dual CFT was proposed in~\cite{Ryu:2006bv, Hubeny:2007xt}. The entanglement entropy of a subregion $A$ in a $d$ dimensional CFT is given by the minimal area of a bulk co-dimension 2 surface $\gamma_{A}$ homologous to $A$ at the boundary $AdS_{d+1}$ geometry, \emph{i.e.}
\begin{equation}\label{RT}
	S_{A}=\underset{\gamma_{A}}{\text{min}}\frac{\text{Area}(\gamma_{A})}{4 G_{N}}\,.
\end{equation}
This concept has attracted enormous attention to probe various phase transitions e.g. in black holes~\cite{Chaturvedi_2016, Li:2017gyc, zeng2016holographic}, holographic superconductors~\cite{Albash:2012pd, Cai:2012sk, Cai:2012nm, Dey_2014}
metal-insulator transitions \cite{Ling_2016,Ling_2017,Li_2019} and topological transitions \cite{Baggioli:2020cld} in the context of the gravity/condensed matter correspondence \cite{Takayanagi:2014rue}. While formula~\eqref{RT} applies for time-independent case only, a covariant generalization of this formula to general time-dependent case was proposed in~\cite{Hubeny:2007xt}. In this respect, the authors of~\cite{Hartman:2013qma} demonstrated that the
extremal surface passes through the interior region of the black brane for early times~\cite{Hartman:2013qma}. Consequently, the growth of the entanglement entropy is related to the growth of the extremal surface along the spacelike surfaces with small curvatures in the interior region of the black brane. At late times, the extremal surface eventually stops expanding on a specific critical surface inside the horizon and does not approach the
singularity. 

Hence, at late times there is a linear growth in the entanglement entropy with time, from which one defines the entanglement velocity $v_{E}$~\cite{Hartman:2013qma}:
\begin{equation}
	\frac{dS}{dt(0)}=v_{E} \mathcal{V}_{1} s, \quad v_{E}^2= r_{+}^4 \frac{|f|e^{-\chi}}{r^4}|_{r=r_{crit}}\,,
\end{equation}
where $t(0)$ is the boundary time, $\mathcal{V}_{1}$ is length of $y$ boundary direction\,\footnote{The bulk surface is fixed at boundary $x=x_0$ and is extended in the boundary $y$ direction, thus in the bulk follows a curve $r(t)$.} and $s$ is thermal entropy density. For our hairy black holes with a Kasner singularity, $-f e^{-\chi}/r^4$ has a maximum inside the horizon at the radius $r_{crit}$.
For the black hole solutions~\eqref{massivebh} and~\eqref{massivebh1}, one can obtain $v_{E}$ at small $\alpha/T$ for which the inner horizon is absent and the interior is similar to the Schwarzschild case with $p_t=-1/3$.
\begin{equation}\label{vebh}
	v_{E}=\Bigg\{ \begin{matrix}
		\frac{\sqrt{3}}{2 \sqrt[3]{2}}- \frac{3 \sqrt{3}}{64 \pi^2}\left(\sqrt[3]{4}-1\right)\left(\frac{\alpha}{T}\right)^2 +\mathcal{O}\left(\frac{\alpha}{T}\right)^4\,, & 
		\hspace{0.5cm}\text{Type I}\\
		\frac{\sqrt{3}}{2 \sqrt[3]{2}}-a_{1}\frac{\sqrt{3}}{32 \sqrt[3]{4} \pi} (2 \sqrt[3]{2}-1)\frac{\alpha}{T}+ \frac{\sqrt{3}}{4096 \pi^2}\left(a_{1}^2 (7-12 \sqrt[3]{2}(1-\sqrt[3]{2})\right)  \\
		-192\,a_{2}(1-\sqrt[3]{4})\left(\frac{\alpha}{T}\right)^2+\mathcal{O}\left(\frac{\alpha}{T}\right)^3\,.&\hspace{0.5cm}\ \text{Type II}
	\end{matrix}
\end{equation} 
\begin{figure}
	\includegraphics[scale=0.3]{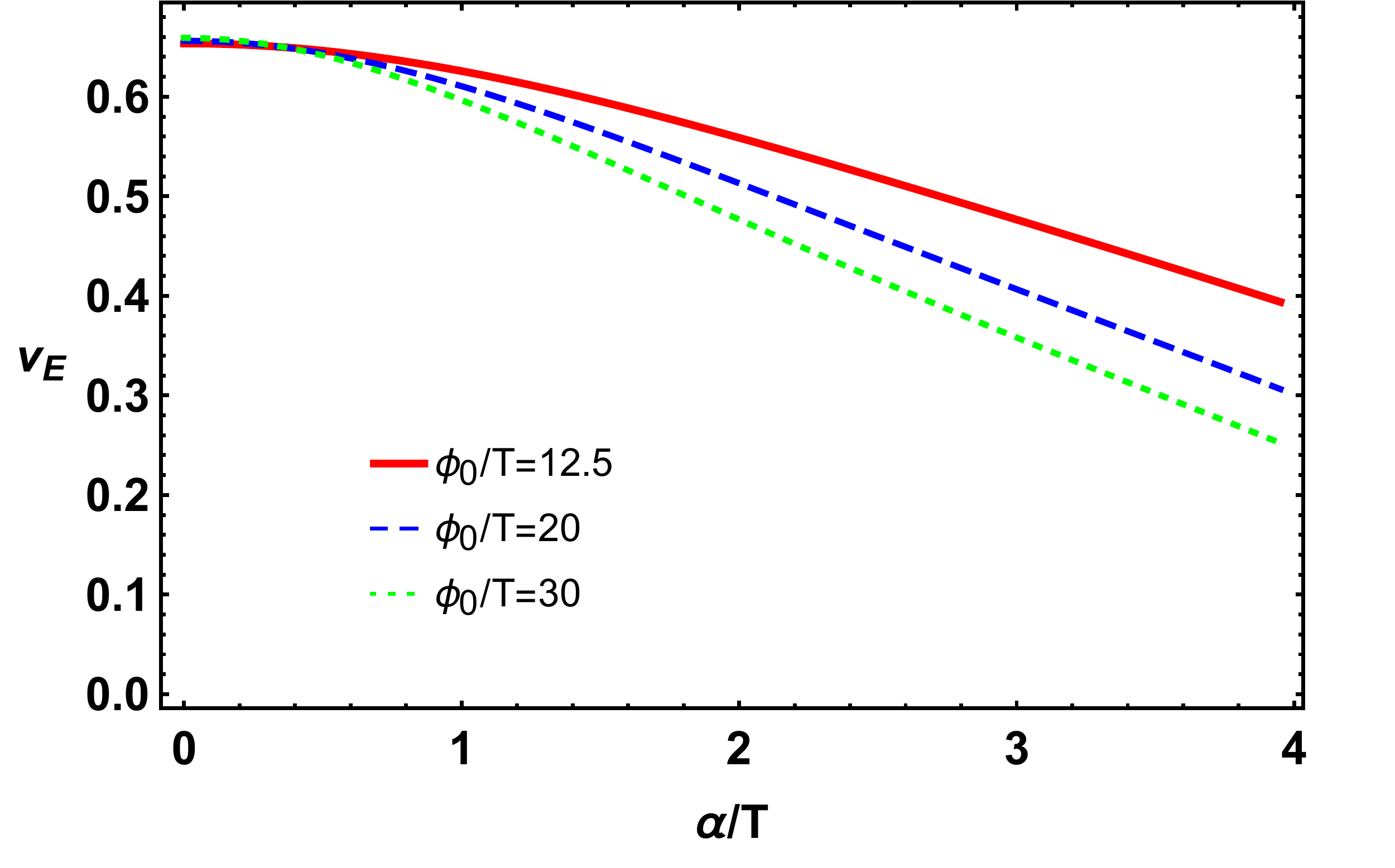}
	\includegraphics[scale=0.3]{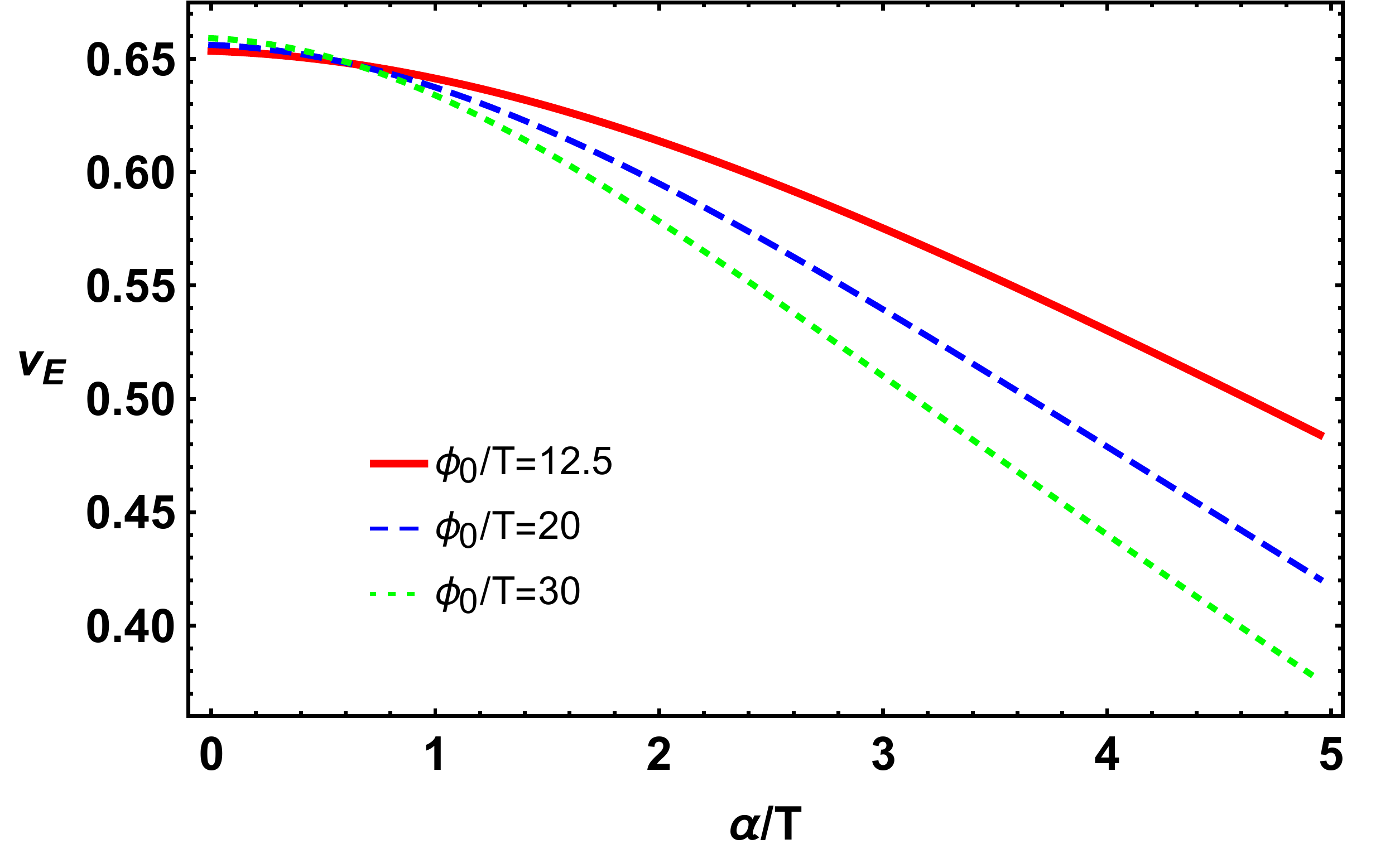}
	\includegraphics[scale=0.3]{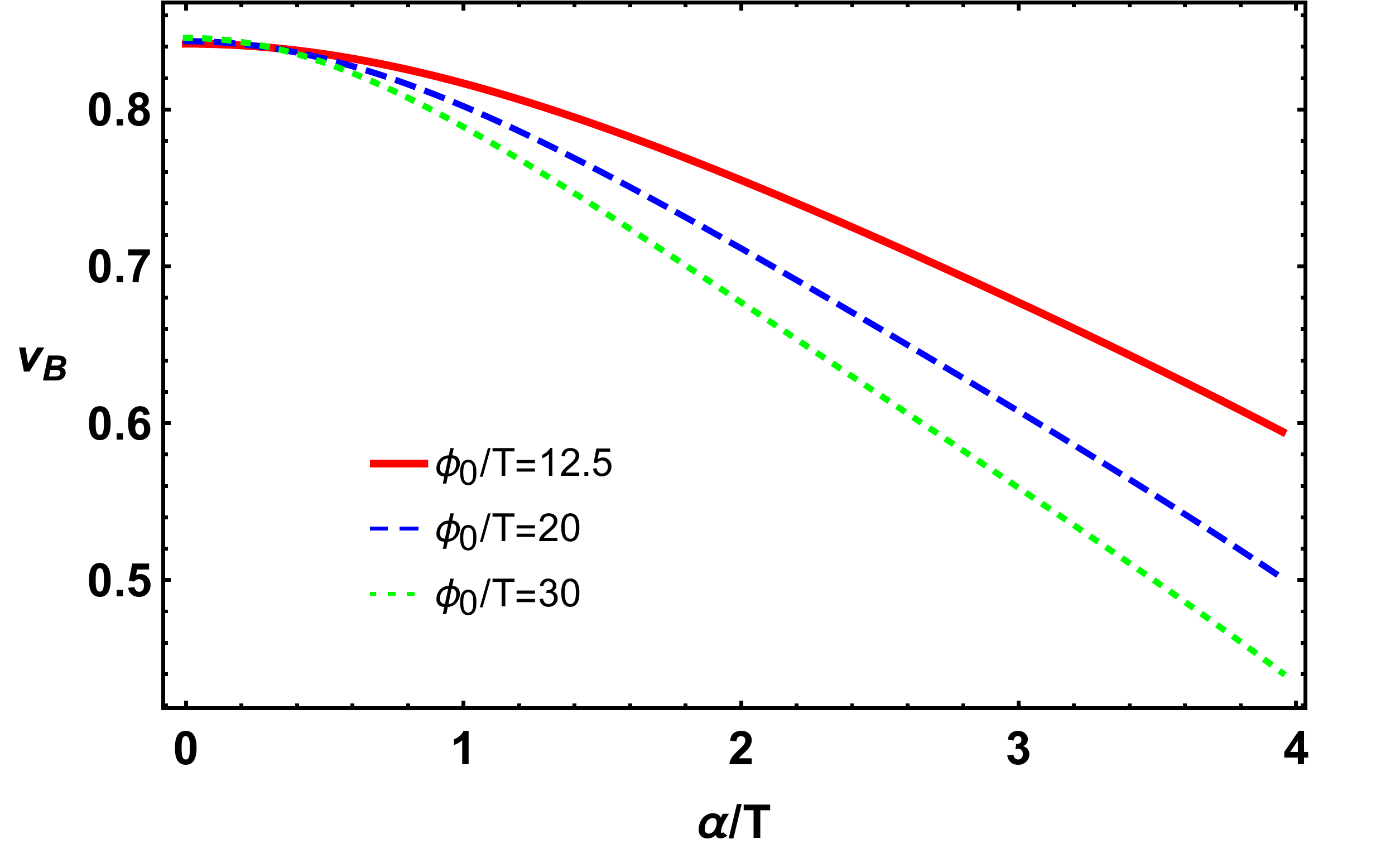}
	\includegraphics[scale=0.3]{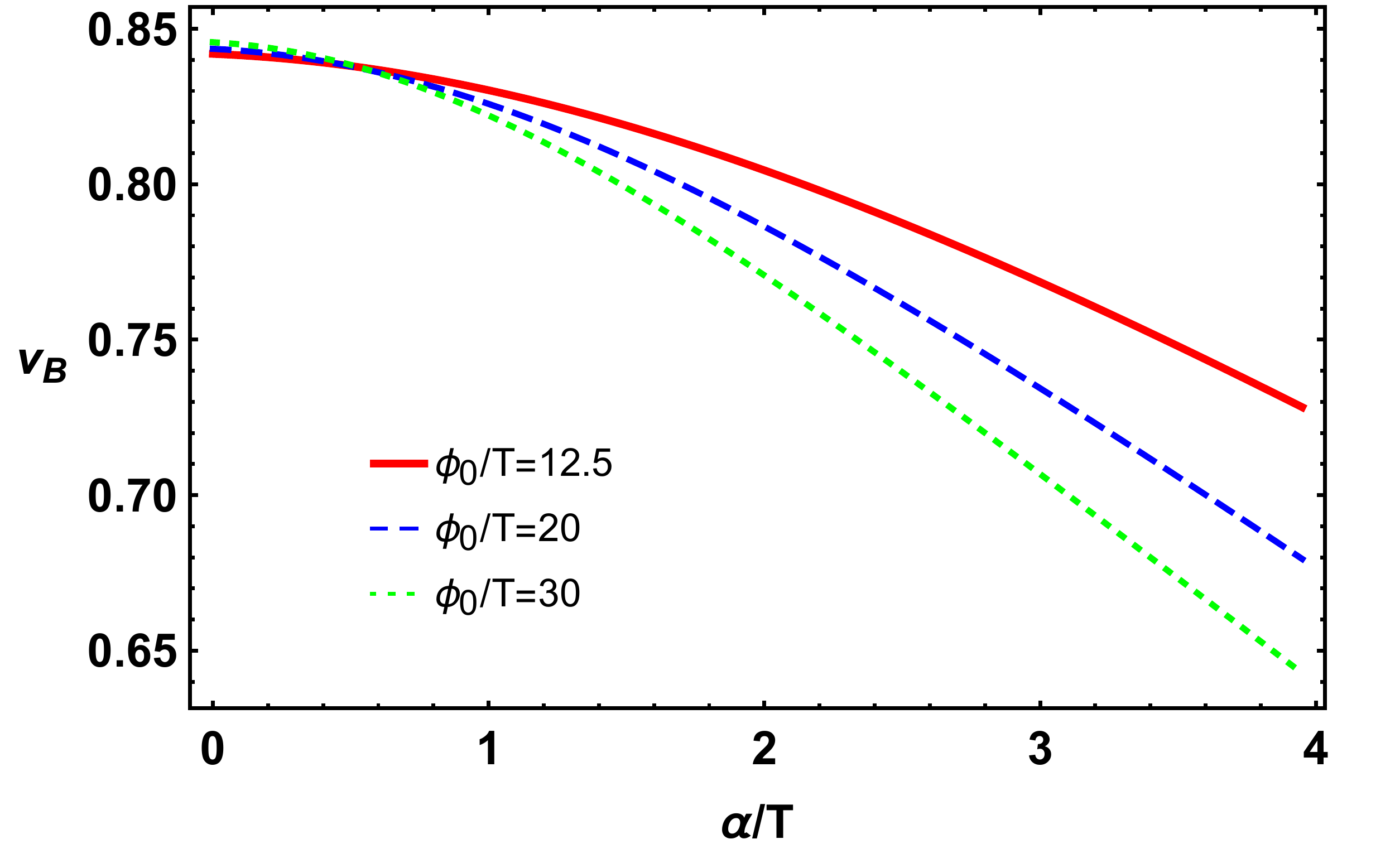}
	\caption{\textbf{Top: }Entanglement velocity $v_{E}$ as a function of $\alpha/T$. \textbf{Bottom:} Butterfly velocity $v_{B}$ as a function of $\alpha/T$.  The \textbf{left panel} is for the Type I case with $K=X$, while the \textbf{right panel} for the Type II case $K=a_{1}\sqrt{X}+a_{2} X$ with $a_{1}=0.1$ and $a_{2}=0.3$.
		\label{fig43}}
\end{figure}

On the other hand, in the context of the gauge/gravity duality, certain properties of quantum chaos in thermal $CFTs$ can be described by the propagation of
shock waves near the event horizon of the $AdS$ black hole \cite{Shenker:2013pqa,Roberts_2015,Leichenauer_2014},
More precisely, the propagation of the
shock wave near the horizon provides a description of
the butterfly effect in the dual field theory. The butterfly velocity in terms of bulk quantities is shown to be~\cite{Shenker:2013pqa}
\begin{equation}
	v_{B}^{2}=r_+ \frac{|f'|}{4} e^{\chi}\Big{|}_{r=r_{+}}\,.
\end{equation}
and it has as well considered as a probe for holographic quantum phase transitions \cite{Baggioli:2018afg}.
For small $\alpha/T$, the butterfly velocity for hairless black hole solutions~\eqref{massivebh} and~\eqref{massivebh1} are, respectively, given by
\begin{equation}\label{vbbh}
	v_{B}= \Bigg\{ \begin{matrix}
		\frac{\sqrt{3}}{2}- \frac{3 \sqrt{3}}{128 \pi^2} \left(\frac{\alpha}{T}\right)^2+\mathcal{O}\left(\frac{\alpha}{T}\right)^4\,, & \hspace{0.5cm}\text{Type I}\\
		\frac{\sqrt{3}}{2}- \frac{\sqrt{3} a_{1}}{32 \pi } \frac{\alpha}{T}+\frac{3 \sqrt{3}}{1024 \pi^2} \left(a_{1}^2-8 a_{2}\right) \left(\frac{\alpha}{T}\right)^2 +\mathcal{O}\left(\frac{\alpha}{T}\right)^3\,, &\hspace{0.5cm}\text{Type II}\\
	\end{matrix}
\end{equation} 

\begin{figure}
	\includegraphics[scale=0.3]{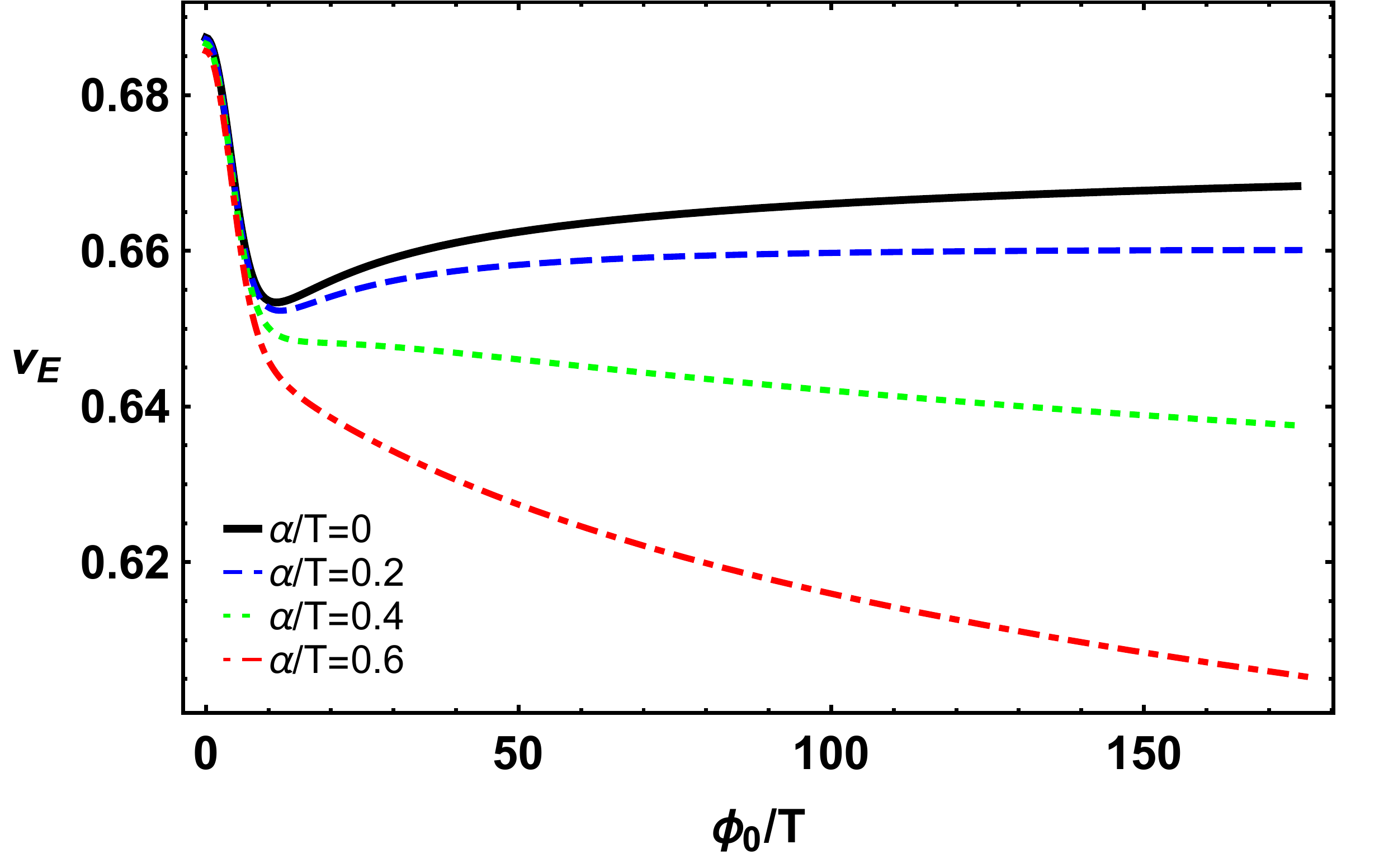}
	\includegraphics[scale=0.3]{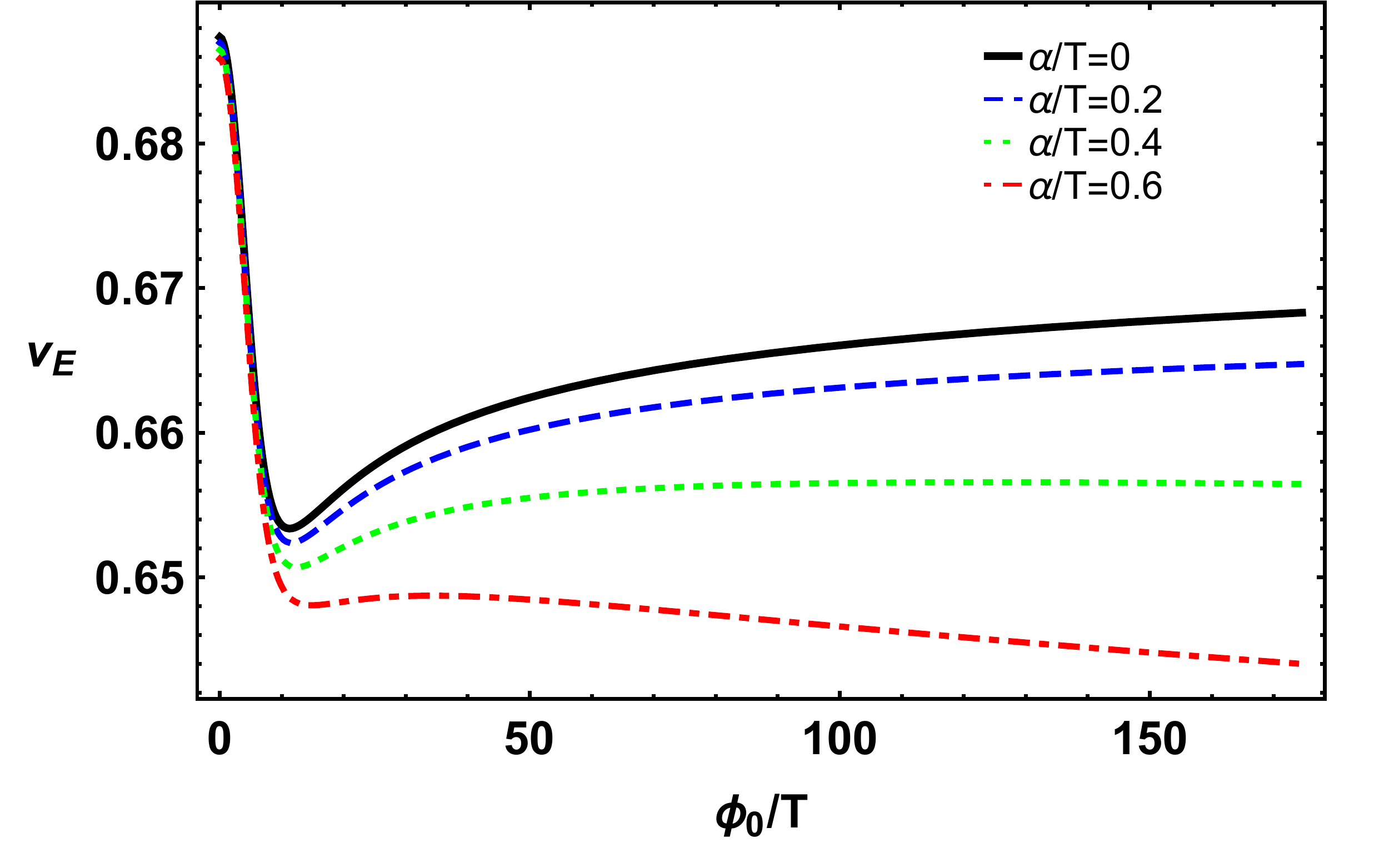}
	\includegraphics[scale=0.3]{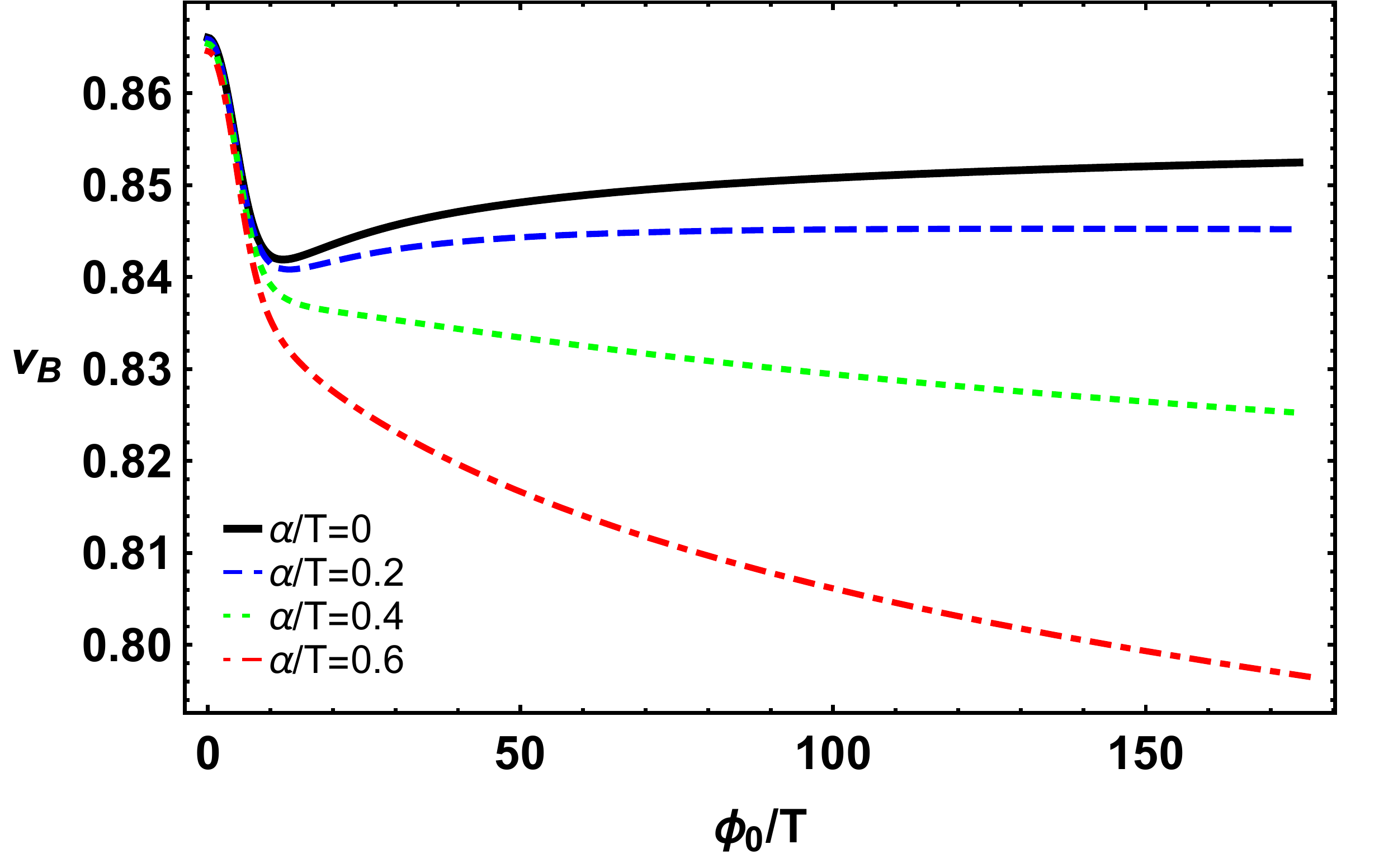}
	\includegraphics[scale=0.3]{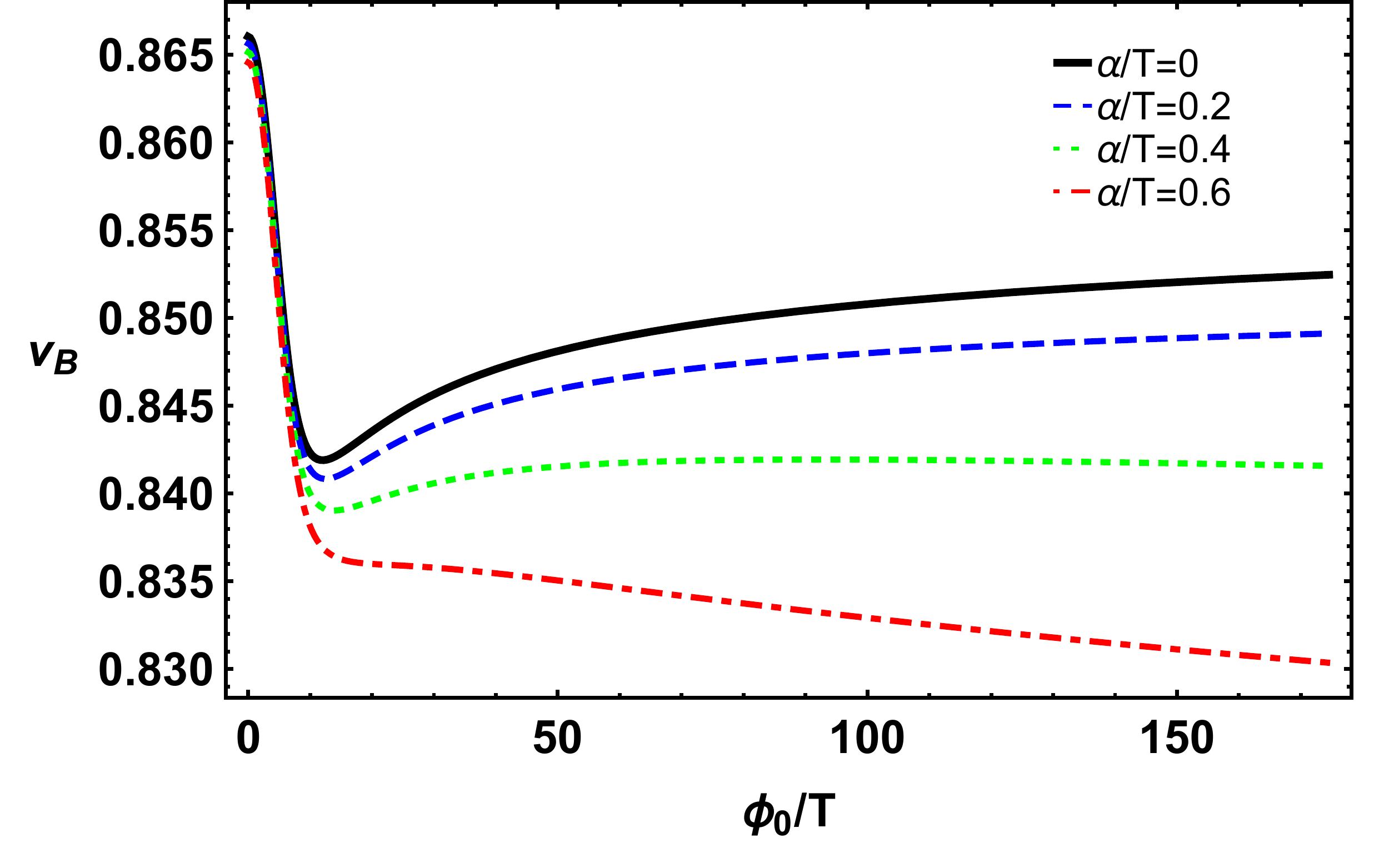}
	\caption{\textbf{Left: }Entanglement velocity $v_{E}$ as a function of $\phi_{0}/T$. \textbf{Right: }the butterfly velocity $v_{B}$ as a function of $\phi_{0}/T$. The \textbf{left panel} corresponds to the Type I case with $K=X$, while the \textbf{right panel} to the Type II case with $K=0.1 \sqrt{X}+0.3 X$.
		\label{fig41}}
\end{figure}

In the presence of the scalar field $\phi$, the black hole interior is deformed. We show the entanglement velocity $v_E$ and the butterfly velocity $v_B$ as a function of $\alpha/T$ in Fig.~\ref{fig43}, and the case as a function of $\phi_0/T$ in Fig.~\ref{fig41}. One can find that $v_E$ and $v_B$ display a similar behavior. Both velocities decrease by increasing $\alpha/T$ for given $\phi_0/T$. For small $\alpha/T$, both $v_E$ and $v_B$ as a function of $\phi_0/T$ first decrease and then increase, while both decrease monotonically  when $\alpha/T$ is large. 

\begin{figure}
	\includegraphics[scale=0.6]{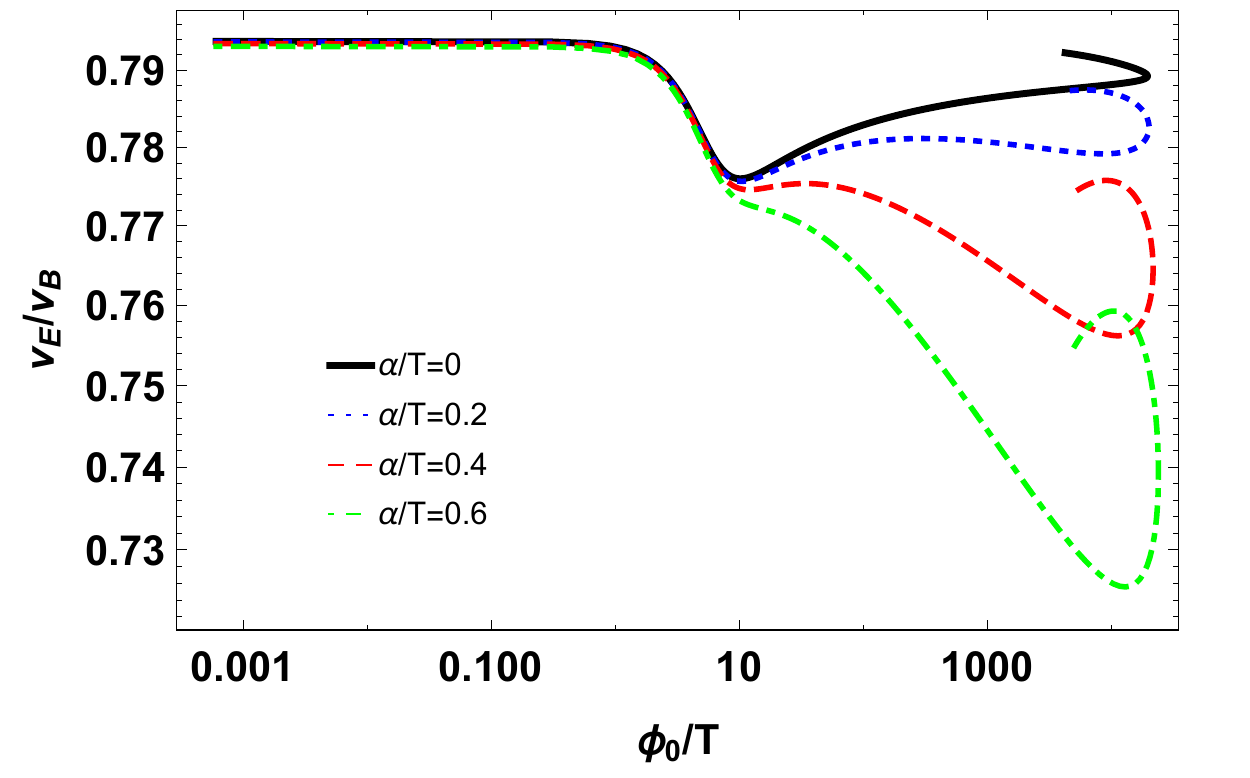}
	\includegraphics[scale=0.6]{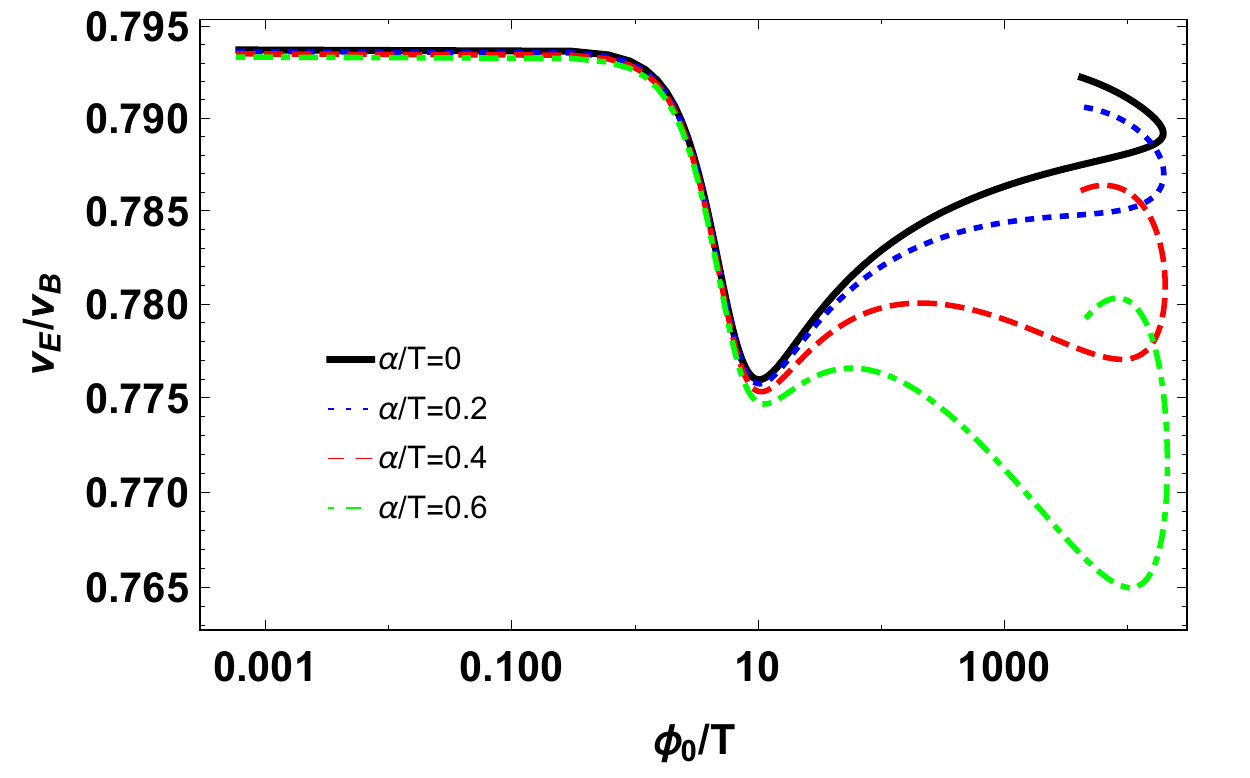}
	\caption{The ratio between $v_{E}$ and $v_{B}$ as a function of $\phi_{0}/T$ by dialing $\alpha/T$. The \textbf{left panel} is for $K=X$, and the \textbf{right panel} for $K=0.1 \sqrt{X}+0.3 X$.
		\label{fig42}}
\end{figure}

\begin{figure}
	\includegraphics[scale=0.3]{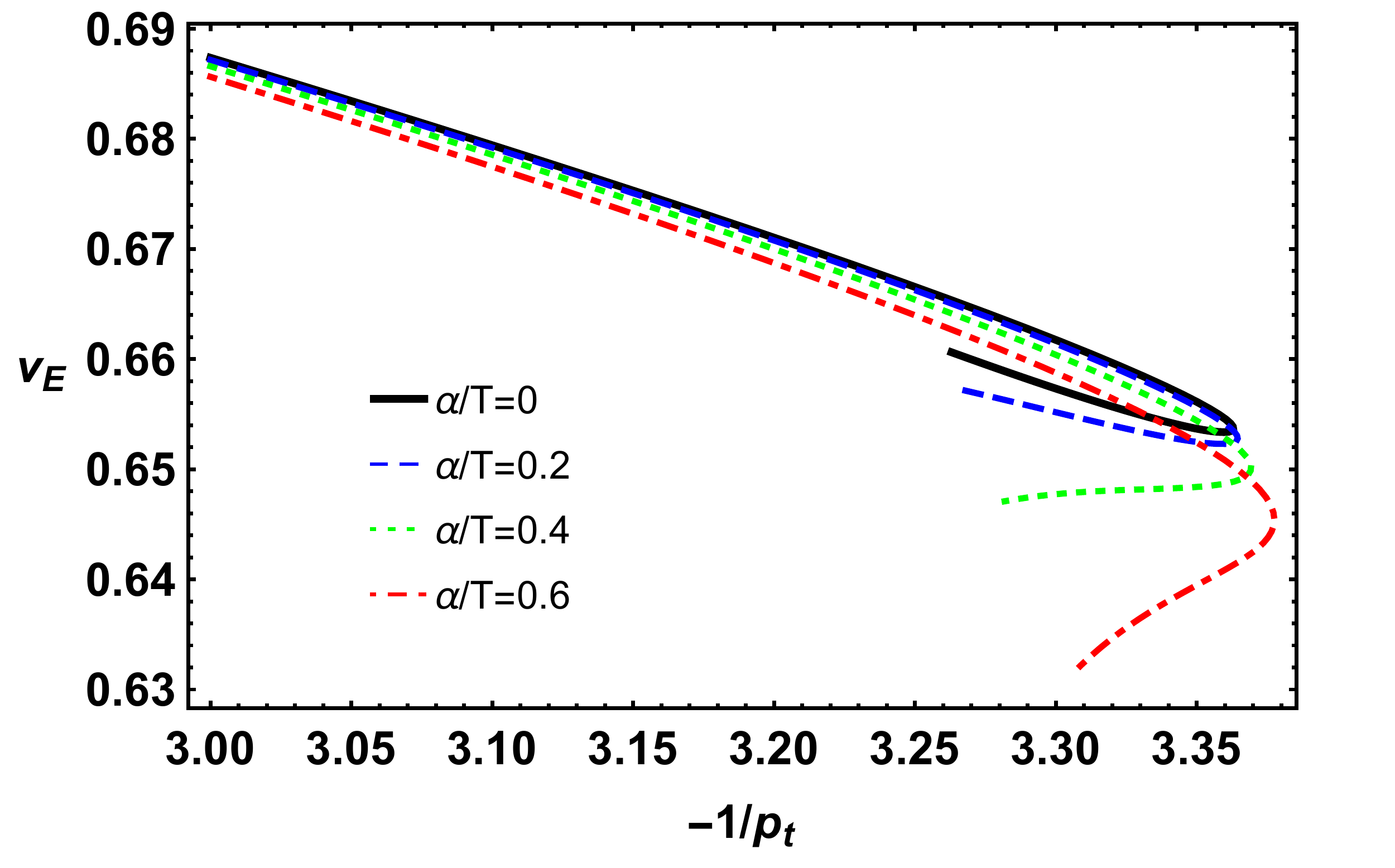}
	\includegraphics[scale=0.3]{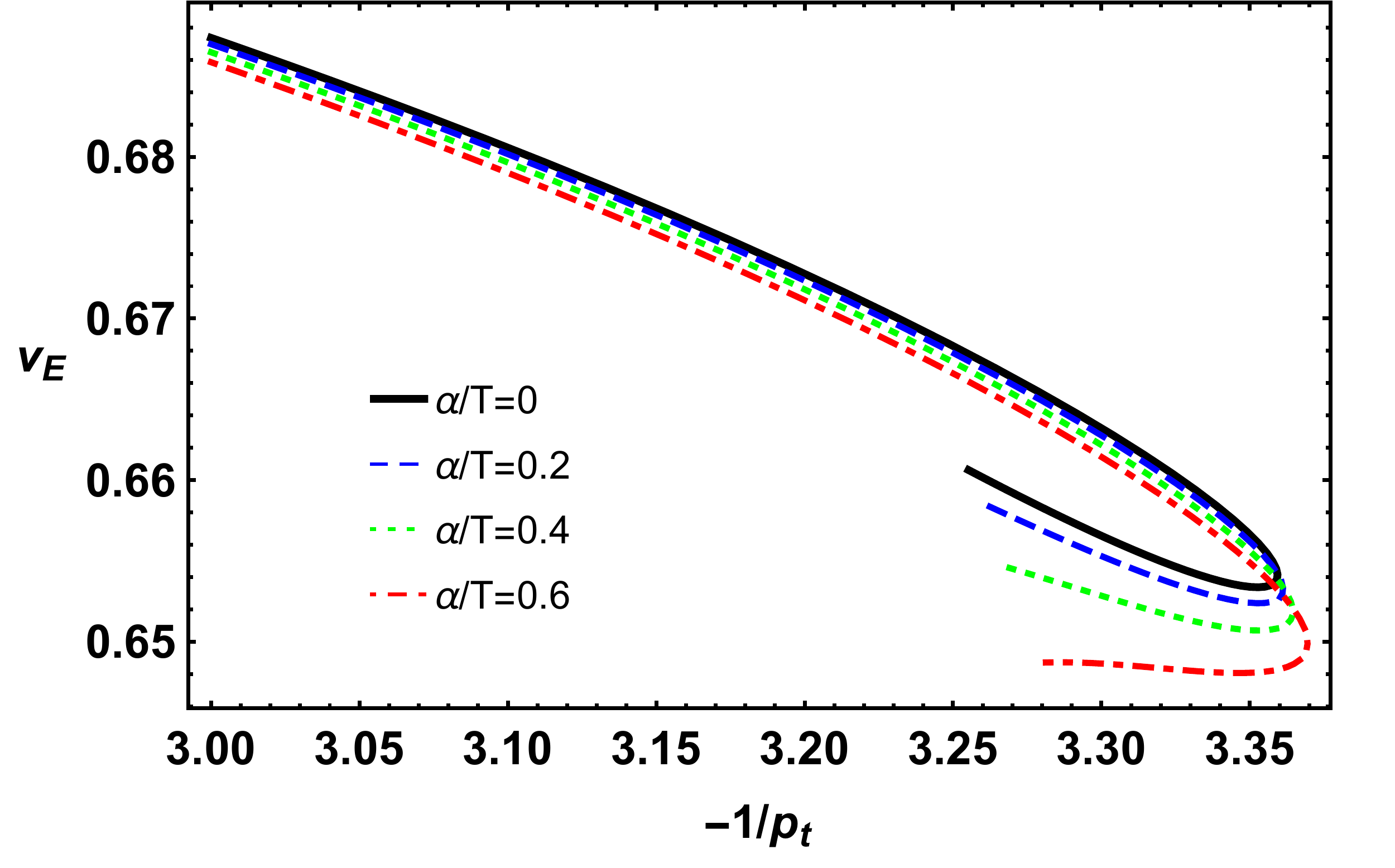}
	\includegraphics[scale=0.3]{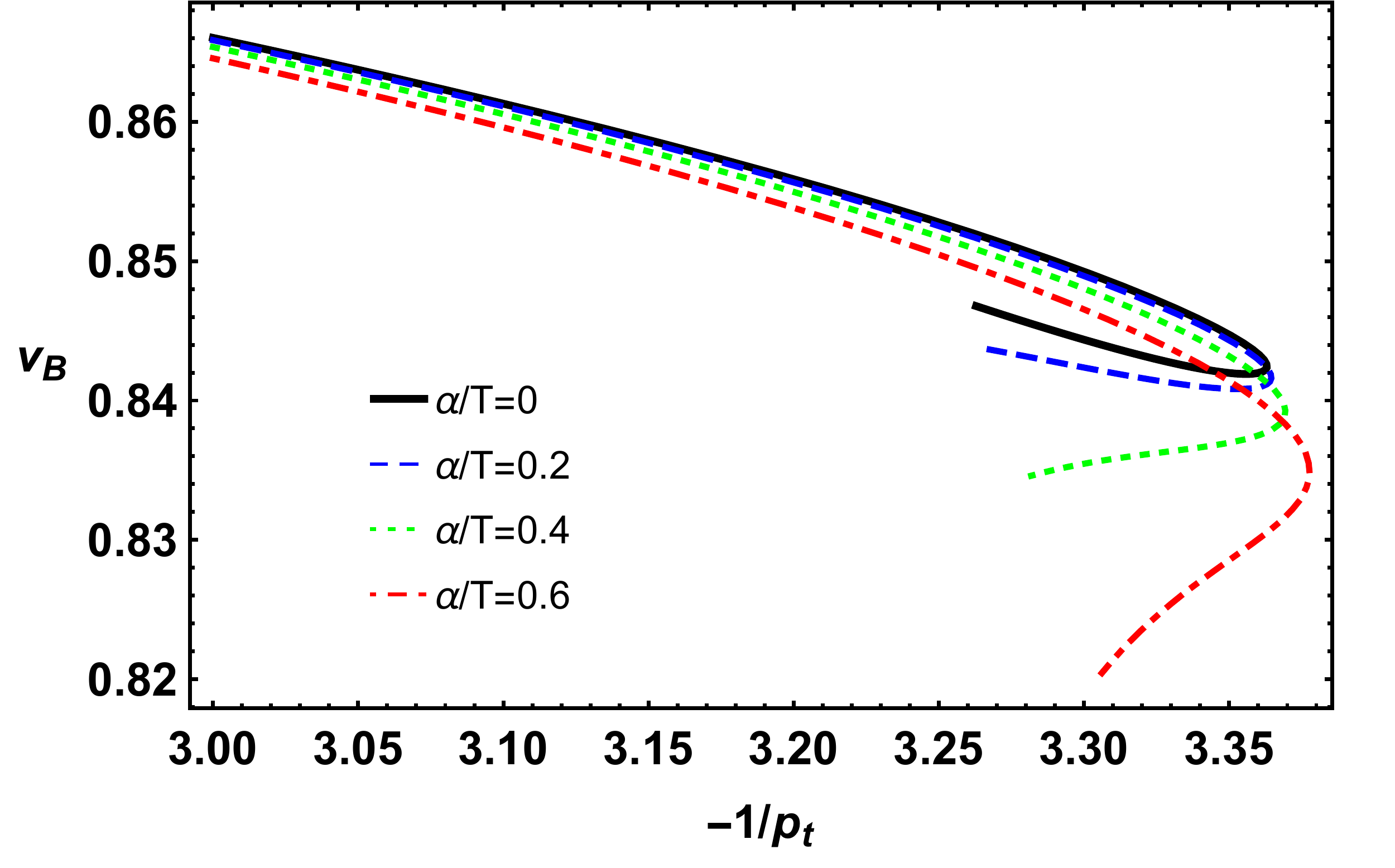}
	\includegraphics[scale=0.3]{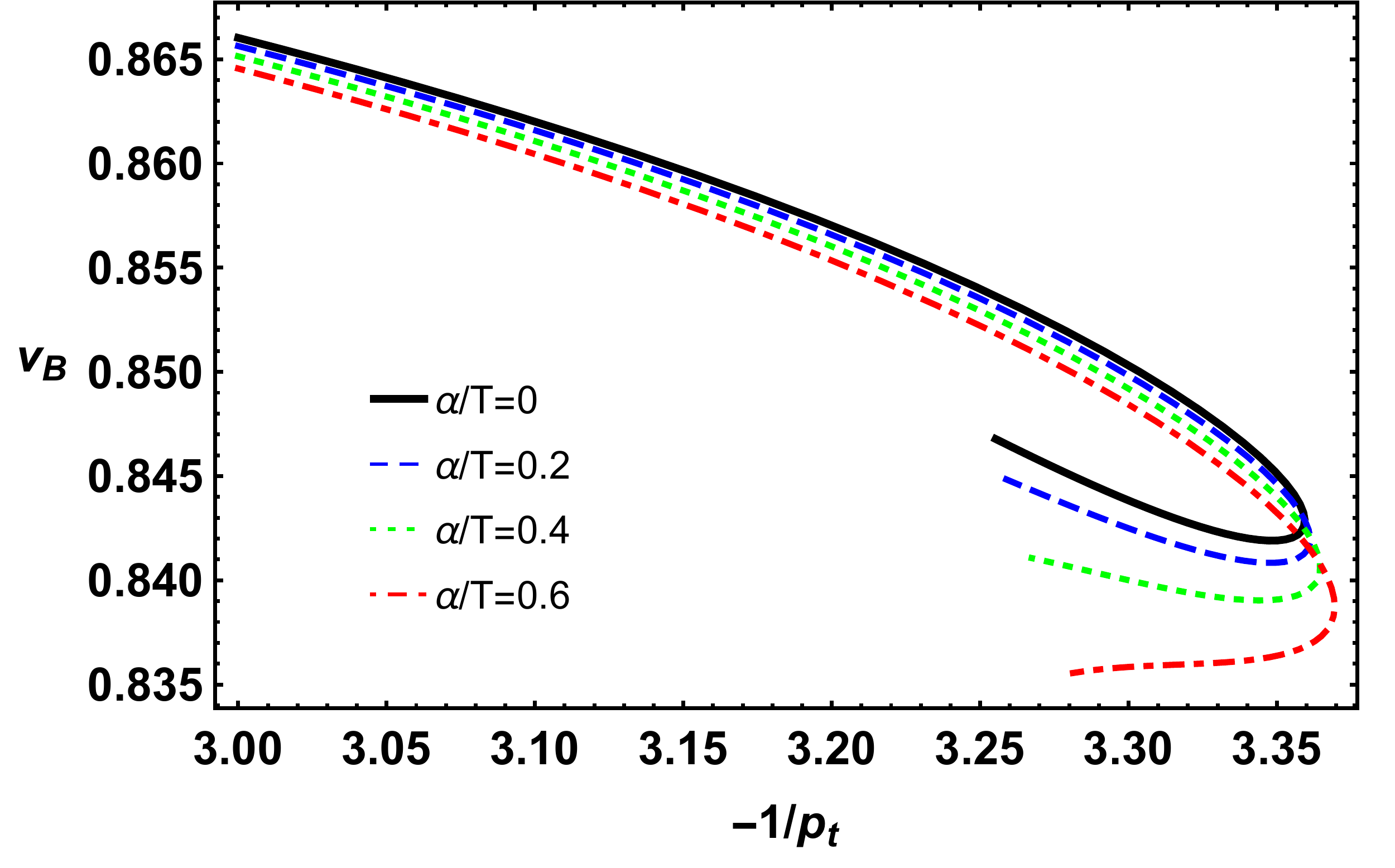}
	\caption{\textbf{Top: } Entanglement velocity $v_{E}$ as a function of $-1/p_{t}$. \textbf{Bottom: } Butterfly velocity $v_{B}$ as a function of $-1/p_{t}$. $K=X$ for the \textbf{left panel} and $K=0.1 \sqrt{X}+0.3 X$ for the \textbf{right panel}.
	\label{fig44}}
\end{figure}

Nevertheless, as shown in Fig.~\ref{fig42}, the value of butterfly velocity is always bigger than the entanglement velocity, consistent with the result proved in~\cite{Mezei:2016zxg}. By setting $\alpha=0$, our model comes back to the one studied in~\cite{Frenkel:2020ysx} where the hairless background is given by Schwarzschild solution. We find that the ratio $v_{E}/v_{B}$ tends to the Schwarzschild value $v_{E}/v_{B}=1/\sqrt[3]{2}$ as $\phi_{0}/T \to \infty$, which is consistent with the expectation in~\cite{Frenkel:2020ysx}. In contrast, for the massive gravity case ,the ratio $v_{E}/v_{B}$ at large $\phi_{0}/T$ is significantly different from the Schwarzschild value $v_{E}/v_{B}=1/\sqrt[3]{2}$. This suggests that $v_E$ and $v_B$ can indeed probe the region behind the event horizon. As a property of the black hole interior, we show $v_E$ and $v_B$ as a function of the Kasner exponent in Fig.~\ref{fig44}. For our solutions with a deformed interior, both velocities decrease away from the undeformed values at $-1/p_t=3$. As $-1/p_t$ is increased, the behaviors are sensitive to the value of $\alpha/T$.

\section{Conclusions}\label{sec:discussion}

We have studied holographic RG flows from 3-dimensional UV CFTs to the Kasner universe in the \color{black} trans-IR \color{black} driven by a relevant scalar operator in massive gravity theories. In the bulk, this scenario corresponds to black hole solutions at finite temperature with a non-trivial scalar hair $\phi$. Due to the mass of graviton, the black hole in the absence of $\phi$ can have an inner Cauchy horizon, and thus has a similar internal structure of the RN black hole. Nevertheless, we have shown that the Cauchy horizon never develops for any scalar potential with $\diff{V}{\phi^2}<0$. Therefore, the hairy solution continues smoothly through the event horizon and ends in a spacelike singularity at later interior time. Moreover, we have shown that the instability of the inner horizon results in the collapse of the Einstein-Rosen bridge, for which $g_{tt}$ rapidly collapses to an exponential small value over a shot proper time. We have found that the spacelike singularity takes a general Kasner form as long as the the potentials terms can be neglected. We have also uncovered that the Kasner form can be violated when the potential terms become important to determine the resulting black hole geometry.  Finally, we have used the entanglement velocity and butterfly velocity to probe of the black hole interior. In particular, we studied in detail the effect of the scalar deformation and the graviton mass on the Kasner exponents.

In the present work, we have limited ourselves to black holes with maximally symmetric horizons; it would be interesting to consider more general cases with inhomogeneous geometries and with additional matter fields, for example, the case of holographic superconductors. Interestingly, we have shown that the dynamics near the spacelike singularity allows for different behaviors with respect to the standard Kasner form, it is worth investigating this feature in the future in more detail. While both entanglement and butterfly velocities seem to be a property of the black hole interior, they are not able to probe the near-singularity region. It will be helpful to find other probes for the Kasner singularity. Finally, it would be desirable to consider quantum corrections to the interior of black holes, in particular, in the vicinity of the singularity. We hope to report on results in the above directions in the near future.

\section*{Acknowledgements}
We would like to thank S. A. Hartnoll for helpful discussions on the numerics. L.L. is supported in part by the National Natural Science Foundation of China Grants No.12075298, No.11991052 and No.12047503, and by the Key Research Program of the Chinese Academy of Sciences (CAS) Grant NO. XDPB15 and the CAS Project for Young Scientists in Basic Research YSBR-006.
M.B. acknowledges the support of the  Shanghai Municipal Science and Technology Major Project (Grant No.2019SHZDZX01).

 \newpage
 
\bibliographystyle{JHEP}
\bibliography{kasner}

\end{document}